\documentclass[a4paper,fleqn,usenatbib]{mnras} 
\usepackage{newtxtext,newtxmath}
\usepackage[T1]{fontenc}
\usepackage{ae,aecompl}

\usepackage{graphicx}	% Including figure files
\usepackage{amsmath}	% Advanced maths commands
\usepackage{multicol}        % Multi-column entries in tables
\usepackage{bm}		% Bold maths symbols, including upright Greek

\newcommand{\dn}{\hbox{$\rm D4000_n$}}
\newcommand{\hb}{\hbox{H$\beta$}}
\newcommand{\hdg}{\hbox{H$\delta_A$+H$\gamma_A$}}
\newcommand{\mgfep}{\hbox{$\rm [MgFe]^\prime$}}

\newcommand{\mgtwofe}{\hbox{$\rm [Mg_2Fe]$}}

\newcommand{\mgtfe}{\hbox{$\rm Mgb/\langle Fe\rangle$}}

\newcommand{\afe}{\hbox{$\rm[\alpha/Fe]$}}
\newcommand{\Dmgfe}{\hbox{$\rm \Delta(Mgb/\langle Fe\rangle)$}}

\def\aj{AJ}%
\def\apj{ApJ}%
\def\apjl{ApJ}%
\def\apjs{ApJS}%
\def\ao{Appl.~Opt.}%
\def\apss{Ap\&SS}%
\def\aap{A\&A}%
\def\mnras{MNRAS}%
\def\pasp{PASP}%
\def\nat{Nature}%
\title[Ages, metallicities and \afe~of centrals and satellites]{Galaxy evolution across environments as probed by the ages, stellar metallicities and \afe~of central and satellite galaxies}

\author[A. R. Gallazzi et al.]{Anna R. Gallazzi$^{1}$\thanks{E-mail:
anna.gallazzi@inaf.it}, A. Pasquali$^{2}$, S. Zibetti$^{1}$, F. La Barbera$^{3}$\\
$^{1}$INAF-Osservatorio Astrofisico di Arcetri, Largo Enrico Fermi 5, 50125 Firenze, 
Italy\\
$^{2}$Astronomisches Rechen-Institut, Zentrum f\"ur Astronomie der Universit\"at Heidelberg, M\"onchhofstr. 12-14, D-69120 Heidelberg, Germany\\
$^{3}$INAF-Osservatorio Astronomico di Capodimonte, sal. Moiariello 16, 80131 Napoli, Italy}

\date{Accepted 2021 January 22. Received 2020 December 23; in original form 2020 October 09}
\pubyear{}

\begin{document}
\label{firstpage}
\pagerange{\pageref{firstpage}--\pageref{lastpage}}

\maketitle

\begin{abstract}
We explore how the star formation and metal enrichment histories of present-day galaxies have been affected by environment combining stellar population parameter estimates and group environment characterization for SDSS DR7.
We compare stellar ages, stellar metallicities and, crucially, element abundance ratios \afe~of satellite and 
central galaxies, as a function of their stellar and host group halo mass, controlling for the current star formation rate and for the infall epoch.
We confirm that below $M_\ast\sim10^{10.5}M_\odot$ satellites are older and slightly metal-richer than equally-massive 
central galaxies. On the contrary, we do not detect any difference in their \afe: \afe~depends primarily on stellar mass and not on group hierarchy 
nor host halo mass. 
We also find that the differences in the median age and metallicity of satellites 
and centrals at stellar mass below $\rm 10^{10.5}M_\odot$ are largely due to the higher fraction of passive galaxies among satellites and as a function of halo mass. We argue that the observed trends at low masses reveal the action of satellite-specific environmental effects in a `delayed-then-rapid' fashion.
When accounting for the varying quiescent fraction, small residual excess in age, metallicity and \afe~emerge for satellites dominated by old stellar populations and residing in halos more massive than $10^{14}M_\odot$, compared to equally-massive central galaxies. This excess in age, metallicity and \afe~pertain to ancient infallers, i.e. satellites that have accreted onto the current halo more than 5 Gyr ago. This result points to the action of environment in the early phases of star formation in galaxies located close to cosmic density peaks.
\end{abstract}

%%%%%%%%%%%%%%%%%%%%%%%%%%%%%%%%%%%%%%%%%%%%%%%%%%

%%%%%%%%%%%%%%%%% BODY OF PAPER %%%%%%%%%%%%%%%%%%

\section{Introduction}

In the local Universe several galaxy properties, including morphology \citep{Dressler80,poggianti08}, star formation rate \citep{Gavazzi02,Gomez03,balogh04,poggianti08}, atomic gas content \citep{GiovanelliHaynes83,chung09,Catinella13}, colors \citep{weinmann06,baldry06} correlate with environmental density. These same properties are also known to correlate with galaxy mass \citep{kauffmann03,Jarle04,Tremonti04,gallazzi05,mcdermid15,Catinella18}. In summary, quiescent galaxies with low gas content and elliptical-like morphologies dominate in terms of number density and mass density the high-mass end of the galaxy mass function \citep{baldry04,baldry12,moffett16} and in high-density environments \citep{baldry06}.
This result and some of the aforementioned correlations with environmental density and with mass have been observed out to $z<1$ \citep{Cooper10b,peng10, Cucciati17}.
It is only thanks to spectroscopic surveys with large statistics that it is possible to disentangle trends with environment and trends with galaxy mass \citep[e.g.][]{baldry06,Cooper08,pasquali10,peng10,Woo13}.

Several processes can affect the amount of gas in galaxies and its ability to form stars eventually leading to suppression of star formation activity and galaxy quenching. These can be broadly distinguished into secular/internal processes \citep[`mass quenching', presence of bulge stabilizing disk, stellar feedback, AGN feedback;][]{Kormendy13,DekelBirnboim06} and environmental processes acting on galaxies that become satellites of groups or clusters \citep[ram-pressure stripping, strangulation, galaxy harassment, tidal interactions, mergers; see for a review][]{BoselliGavazzi06}. 
Distinguishing the action of environment-specific processes from that of internal processes is complicated by the fact that the environment in which galaxies reside evolves with redshift and so also the environment-specific processes that a galaxy experience vary in time and in efficiency. 
Moreover, environment can influence the growth history of galaxies both through satellite's specific processes that act on galaxies as they enter a halo (`nurture') and through setting different initial conditions on the efficiency of star formation, feedback processes and mergers depending on the location of galaxies in the cosmic density field (`nature') \citep{DeLucia2007}.

While it is not straightforward to define which `environment' is most relevant in galaxy evolution \citep{muldrew12,Wilman10}, a convenient way to try and isolate the effects of mechanisms acting specifically on galaxies falling into groups or clusters is by distinguishing galaxies into `centrals' and `satellites'. This is a natural scheme in the framework of hierarchical structure growth of hydrodynamical simulations and of semi-analytic models. Though not free from uncertainties and potential misclassifications especially in the regime of small groups \citep[e.g.][]{delucia14}, it has proven to be a powerful framework to be applied to large spectroscopic surveys \citep[e.g.][]{vdB08} allowing a more direct comparison with hierarchical models and simulations of galaxy formation, and allowing to control for both the stellar mass of galaxies and the halo mass of the host environment.

It is now established that the fraction of quiescent galaxies is a function not only of mass, but also of galaxy hierarchy and host halo mass \citep{baldry06,peng10,wetzel12}. This result has provided an important observational testbed for galaxy and structure formation models and their adopted schemes for the efficiency and timescales of quenching mechanisms, both internal and external \citep[e.g.][]{weinmann06b,hirschmann14,bahe17b}. While the abundance of quiescent galaxies increases with both stellar mass and halo mass, the distribution in galaxy colors and star formation rate is bimodal in all environments with little variation of the location of the peaks \citep{baldry06,McGee11,wetzel12}. Combining these two results and infall histories from N-body simulations, \cite{wetzel13} put forward the so-called `delayed-then-rapid' quenching, whereby it takes a considerable time (few Gyr) when a galaxy first become a satellite before rapid quenching occurs. 

More detailed characterization of galaxy physical properties as a function of both mass and environment is needed to gain further insight into when and where galaxies quench under the action of environment. In particular, the stellar population properties are fossil records of the past star formation and metal enrichment history. The luminosity-weighted mean age of the stellar populations reflects the main epoch of mass build-up, modulated by the occurrence of recent (or ongoing) star formation even if at low levels. In \cite{pasquali10} the differences in light-weighted age between centrals and satellites, and as a function of stellar mass and halo mass, provided evidence for an earlier epoch of quenching of satellites in more massive halos with respect to equally-massive centrals. The observed trends were understood in the context of the semi-analytic model of \cite{Wang08} by removal of the hot halo gas in satellites upon infall and by the earlier infall epochs of satellites in today's massive halos. 

The stellar metallicity, integrated over the whole star formation history, reflects the efficiency of star formation, metal production and metal recycling. Mechanisms that alter the gas content and the chemical properties of the interstellar medium (ISM) will also affect the stellar metallicity. Hence differences in stellar metallicity can be informative not only of any variation in the efficiency of star formation but also of modifications of the ISM chemical properties. For the general galaxy population, the different stellar metallicity-mass relations for star-forming and passive galaxies have been shown to provide important constraints on the main quenching mechanism \citep{peng15,Trussler20a} and on inflow/outflow efficiency \citep{spitoni17}. In \cite{pasquali10} we showed that satellites are metal-richer than equally-massive centrals at masses below $10^{10.5}M_\odot$ with the difference being larger in higher-mass halos. When restricting the sample to star-forming galaxies in \cite{pasquali12}, the difference in stellar metallicity between centrals and satellites at fixed mass vanishes, but it emerges an excess in the gas-phase metallicity of satellites compared to centrals. The observed trends in stellar and gas-phase metallicities favor mechanisms that deprive galaxies of their gas content (strangulation and then ram-pressure stripping) inhibiting gas inflows leading to enhanced metal enrichment \citep{pasquali12,bahe17,Maier19a,Trussler20a}.

The element abundance ratio \afe~in stars also represents a mass-weighted property integrated over the whole SFH. The \afe~arises from the relative effective yields of SNII and SNIa products \citep{greggio83}. Several factors can affect this ratio, including variations in IMF, delay-time-distribution of SNIa, differential loss of metals, star-formation timescales \citep{Tinsley79,Matteucci94,trager2000}. However it has become generally accepted, in the assumption of a universal IMF, to interpret the $\afe-M_\ast$ relation of early-type galaxies as indicative of shorter formation timescales in more massive galaxies as a consequence of star formation being interrupted before the Fe-peak elements of SNIa have time to be recycled in stars \citep[e.g.][]{Thomas05}. 
The high level of $\alpha$-enhancement of massive early-type/quiescent galaxies and its relation with stellar mass \citep[e.g.][]{trager2000,kuntschner01,graves09,mcdermid15} are a strong testbed for cosmological simulations of galaxy formation and semi-analytic models. The observed \afe~of quiescent galaxies can be reproduced by requiring a top-heavy IMF during intense star formation events \citep{nagashima05,fontanot17} or very short timescales achieved through the action of AGN feedback \citep{pipino09,segers16}, possibly in addition to starbursts triggered by fly-by encounters \citep{CaluraMenci11}. \cite{segers16} have shown that EAGLE cosmological hydrodynamical simulations that reproduce the main scaling relations observed for local early-type galaxies \citep[including the mass-metallicity relation -][]{schaye15} can also reproduce the trend between \afe~and stellar mass for masses above $\rm 10^{10.5}M_\odot$ as a result of AGN feedback quenching star formation in massive galaxies. However, \cite{delucia17} discuss, in the framework of the semi-analytic GAEA model, that quenching induced by AGN feedback is not enough to reach the observed levels of \afe. Imposing a truncation of star-formation in galaxies more massive than $\rm 10^{10.5}M_\odot$ appears necessary to reproduce the \afe-M$_\ast$ relation, but would violate the stellar mass-metallicity relation. This result led \cite{fontanot17} to propose an alternative interpretation, by which  \afe~is not trivially related to star formation timescale, but rather to the SFR at the peak of the galaxy star formation history, in the context of a SFR-dependent IMF \citep[IGIMF -][]{WeidnerKroupa05}.

Differences in \afe~as a function of environment can thus be informative of either a shorter formation timescale or different star formation conditions at the peak of activity in galaxies in dense environments. Previous works on early-type galaxies found that the main environmental effect was an increased scatter toward slightly younger ages, lower metallicities and lower \afe~in less dense environments, but that the scaling relations for the bulk of the population were not dependent on the environmental density \citep{gallazzi06,Thomas10,labarbera14}.
Other element abundance ratios, such as [CN/Fe] in addition to [Mg/Fe], have been used as `chemical clocks' to detect small differences in star formation timescales in massive cluster ellipticals \citep{Carretero07}.

In this paper, we aim at inferring how the past star formation and metal enrichment history of galaxies, as summarized by their present-day stellar population properties, are affected by environment. We make use of our own stellar population catalog of light-weighted ages and stellar metallicities for SDSS DR7 derived from a Bayesian analysis of key absorption features as in \cite{gallazzi05}. By combining the stellar population catalog with the SDSS DR7 group catalog of \cite{wang14}, we characterize galaxy `environment' first of all by distinguishing galaxies in terms of group hierarchy, specifically central galaxies and satellite galaxies. We then control for both stellar mass and host halo mass, by looking at trends for centrals and satellites at fixed stellar mass and at fixed halo mass. We build upon our previous work in \cite{pasquali10}, but we make three important additions to our analysis:

{\it i)} With respect to our previous work, we critically add information on the \afe, derived for the first time for galaxies with any star formation activity, with the aim of constraining any difference in star formation timescale induced by environment. 

{\it ii)} In addition to looking at stellar population trends for the population as a whole, we also control for the effect of the varying fraction of quiescent galaxies, by looking at detailed differences as a function of environment for galaxies with similar current specific star formation rate. 

{\it iii)} Since several works have recently indicated overall long timescale for environmental processes to be effective \citep[e.g.][]{delucia12,oman16,hirschmann14} and because the environment in which galaxies live changes with time, we also wish to control for the epoch of infall, i.e. when a galaxy became satellite of its current environment. The location of galaxies in a diagram combining cluster-centric velocity with cluster-centric distance (the phase-space diagram) has been shown to be a promising tool to infer mean infall times of satellites, even with observed projected quantities \citep[e.g.][and references therein]{oman16,pasquali19}. We thus further characterize the history of satellite galaxies by controlling for their infall epoch using the phase-space parametrization of \cite{pasquali19} and \cite{smith19}. 

Combining all these ingredients allows us to put constraints on the most likely mechanisms and timescale for environmental quenching and to detect the action of environment in the early phases of galaxy formation. \cite{Trussler20b} have recently presented a similar analysis on the differences in age and stellar metallicity between centrals and satellites distinguished into star-forming, green valley and passive. With respect to their analysis we take the further step, as discussed above, to add information about \afe~and the epoch of infall.

We describe our sample, the group catalog and the stellar population parameter estimates, including the derivation of the new \afe~estimates, in Sec.~\ref{sec:data}. We explore the scaling relations for the general populations of centrals and satellites in Sec.~\ref{sec:rel_mstar}, and how they depend on host halo mass in Sec.~\ref{sec:rel_mhalo}. In Sec.~\ref{sec:rel_SFR} we quantify and discuss the residual differences between centrals and satellites after removing the dependence of the quiescent fraction on stellar and halo mass. We then discuss our results in light of the infall epoch, distinguishing `ancient' and `recent' infaller satellites based on their phase-space location, in Sec.~\ref{sec:zones}. Finally, we summarize and discuss our findings in Sec.~\ref{sec:conclusions}. Throughout the work, we assume a flat $\Lambda$CDM cosmology with $\Omega_m=0.275$ and $\Omega_\Lambda=0.725$ \citep{WMPA7}, a \cite{Chabrier03} IMF, and $Z_\odot=0.02$ for solar metallicity.

\section{Data}\label{sec:data}
\subsection{The group catalog}\label{sec:cat}
We use the group catalog for SDSS DR7 constructed by \cite{wang14} following \cite{yang07} by applying the halo-based group finder algorithm of
\cite{yang05} to the New York University Value-Added Galaxy Catalog for SDSS DR7 \citep{sdssdr7}. The catalog includes galaxies with an
apparent magnitude (corrected for Galactic extinction) brighter than $r=18$~mag, with redshift $0.01\leq z\leq0.2$~and with redshift
completeness $C_z>0.7$. In this paper we use the version of the group catalog that includes galaxies with SDSS photometry but
redshift from other spectroscopic surveys in addition to galaxies with SDSS spectroscopic redshift, for a total
number of 596851 galaxies. Galaxy magnitudes and colors are based on Petrosian magnitudes, corrected for Galactic extinction and
K+e--corrected to $z=0.1$~following \cite{blanton03}. Note that for galaxies with $r$-band concentration parameter $C>2.6$ Petrosian
magnitudes have been corrected by $-0.1$~mag. Galaxy stellar masses are computed using the color-M/L relation of \cite{bell03}.

For each group the catalog provides estimates of the dark matter halo mass. As in our previous works we use the halo masses based on the
ranking of the total stellar mass, which has been shown by \cite{more11} to be a better predictor of halo mass than the total luminosity.
This estimate is available for groups more massive than $M_h\sim10^{12}h^{-1}M_\odot$ and with one or more members brighter than
$M_r^{0.1}-5\log h=-19.5$~mag. For smaller groups down to $M_h\sim10^{11}h^{-1}M_\odot$, the halo mass provided in the catalog is
estimated from the relationship of \cite{yang08} between the stellar mass of central galaxies and the halo mass. Uncertainties on halo
masses range from $\sim0.35$~dex around $M_h\sim10^{13.5}-10^{14}h^{-1}M_\odot$ to $\sim0.2$~dex at lower and higher masses, according to
tests on mock catalogs in \cite{yang07}.

\begin{figure}
\centerline{\includegraphics[width=9truecm]{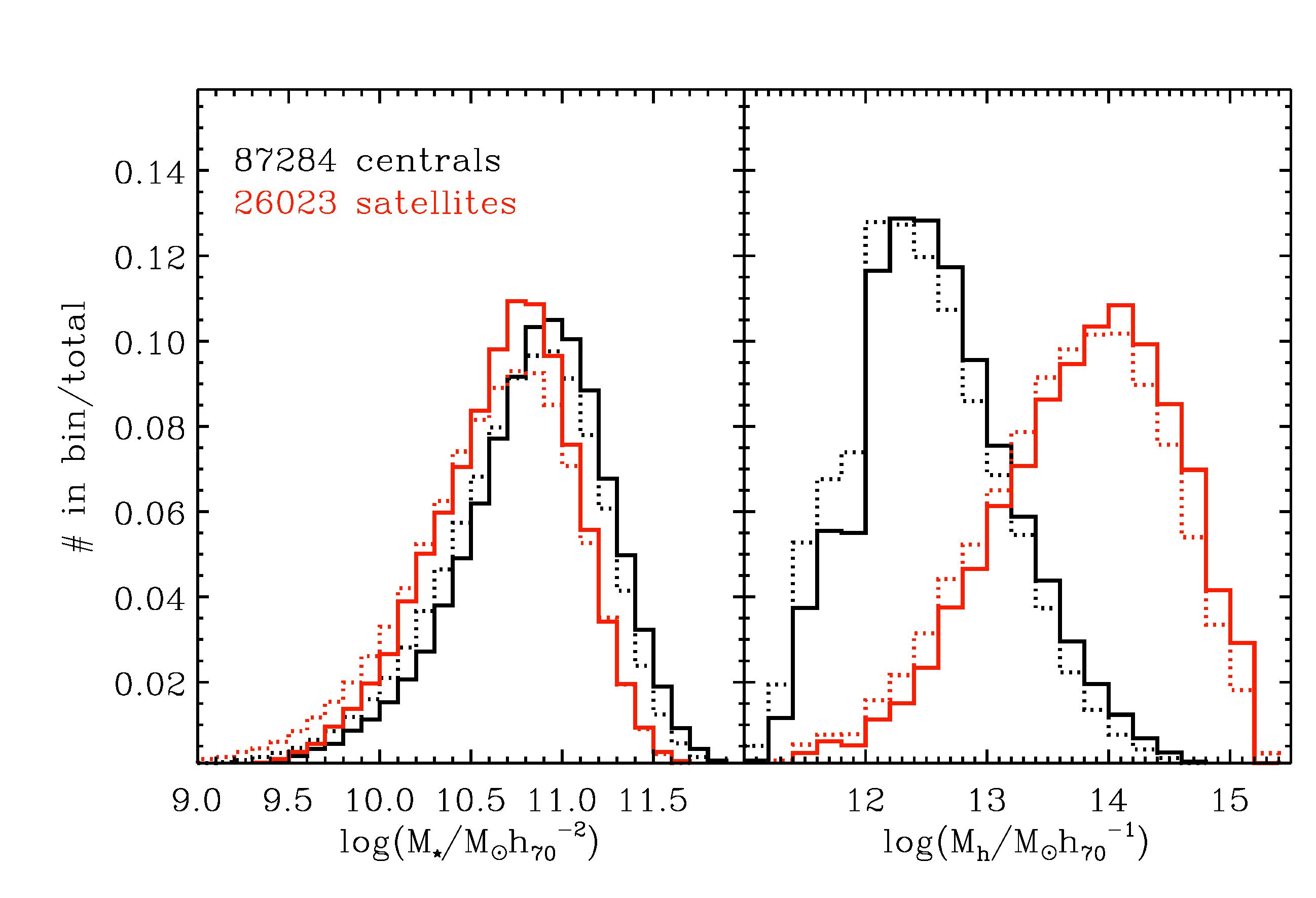}}
\caption{Distribution in galaxy stellar mass ({\it left}) and host group halo mass ({\it right}) for the galaxy sample used in this work.
Galaxies are separated into satellites (red histograms) and centrals (black histograms). The solid histograms show the final subsamples, whose number of galaxies is written in the left panel, while the
dotted histograms show the corresponding distributions for the original sample before the cut at $S/N>20$. 
All histograms are normalised to unity.}\label{fig:mass_distr}
\end{figure}

Galaxies are classified as central galaxies if they are the most massive galaxy in a group or as satellite galaxies otherwise. The catalog comprises 468822 central galaxies and 128029 satellite galaxies. In Fig.\ref{fig:mass_distr}~the dotted histograms show the distribution in stellar mass M$_\ast$ (left) and group halo mass M$_h$ (right) for the two subsamples in the whole group catalog. We probe roughly three orders of magnitude in stellar mass, from $\log (M_\ast/h^{-2}M_\odot)\sim9$~to $\sim12$. As expected, the distribution in stellar mass for central galaxies is shifted to higher stellar masses with respect to satellites. We probe roughly four orders of magnitude in group halo mass, from $\log (M_h/h^{-1}M_\odot)\sim11$~to $\sim15$. The distribution for central galaxies, which reflects the abundance of groups, peaks at $\sim10^{12.5} h^{-1}M_\odot$. The halo mass distribution for satellite galaxies peaks around $\sim10^{14} h^{-1}M_\odot$, a higher value than for central galaxies, as a consequence of the fact that higher-mass groups host a larger number of satellites.

\subsection{Stellar masses}\label{sec:mstar}
Galaxy stellar masses in the \cite{yang07} catalog are computed using the color-M/L relation of \cite{bell03}, specifically the one based on the $g-r$ color and $r$-band magnitude K+e-corrected to $z=0$. In this work we continue to use these stellar mass estimates for consistency with our previous works and with the halo mass estimates in the \cite{yang07} catalog (see Sec.~\ref{sec:cat}). We have compared these stellar masses with those obtained from more up-to-date color-M/L relations that take in better account dust attenuation and complex star formation histories of young galaxies, in particular the updated version of the \cite{zibetti09} $(g-i)-M/L_i$~relation as used in \cite{fontanot17}. The two mass estimates compare generally well with an rms of $0.07$~dex, but with an average offset of $0.15$~dex, and ranging from 0.1~dex at $\log M_\ast=11.5$ to 0.3~dex at $\log M_\ast=9.5$, in the sense that the masses in the catalog are larger. We have checked that we obtain quantitatively consistent results on the scaling relations presented in this work if we use stellar masses computed with the updated version of the \cite{zibetti09} calibration.

We have also compared with the stellar masses estimated from absorption indices in combination with $r-i$ color as in \cite{gallazzi05}. In this case there is a smaller systematic offset ($0.06$~dex in the sense of larger spectroscopic stellar masses) but a larger scatter of $0.13$~dex in particular at intermediate masses. This may be related to a combination of the different treatment of dust attenuation and of aperture effects on the spectroscopic M/L in disk galaxies with prominent bulges. Also in this case we have checked that we find consistent scaling relations for the population as a whole and split according to galaxy hierarchy and group halo mass, presented in Sec.~\ref{sec:rel_mstar} and~\ref{sec:rel_mhalo}. 

\subsection{Ages and stellar metallicities}\label{sec:age_met}
We combine the SDSS group catalog with stellar populations parameters estimated for SDSS DR7 galaxies following \cite{gallazzi05}
(hereafter G05).
Specifically, probability distribution functions (PDF) of r-band luminosity-weighted mean age and stellar metallicities are derived comparing the strength of the \dn, \hb, \hdg,
\mgfep~and \mgtwofe~absorption features to model spectra obtained convolving \cite{bc03} (BC03) SSP models with a large Monte Carlo library of 
star-formation histories and metallicities \footnote{We make use of the catalogs of spectro-photometric data and absorption index measurements, corrected for contamination by sky emission, kindly made publicly available by Jarle Brinchmann at the link http://wwwmpa.mpa-garching.mpg.de/SDSS/DR7/raw\_data.html. The absorption indices are measured from pure stellar continuum galaxy spectra. The removal of emission line contamination is achieved fitting first the emission-line-free regions of each spectrum with a non-negative least-square fit of SSP models and then the residuals are fitted with gaussian-broadened emission line templates \citep[see G05 and][for a more detailed description]{Tremonti04}. With respect to the values provided in the catalog, we increase the error on absorption indices by a factor that accounts for differences between duplicate observations. The error on \dn~is further increased by a 7\% added in quadrature to account for spectro-photometric calibration uncertainties.}. These features have been chosen for their different sensitivity to age and metallicity and
their minimal dependence on element abundance ratios. Notice that BC03 models follow the Galactic abundance pattern, i.e. they are "base" models, approximately scaled-solar at solar metallicity (see Sec.~\ref{sec:afe}). We note that a few works \citep{Vazdekis15, korn05,thomas04} have shown that 4000\AA-break and the high-order Balmer lines \citep[and possibly H$\beta$ -][]{cervantes09} depend on \afe, especially at \dn>1.6. These works agree on the negligible variation of composite Mg-Fe indices with \afe. In G05 we checked that the derived ages and metallicities were not systematically different if we excluded the high-order Balmer lines from the fit. We estimated a possible systematic uncertainty of 0.05~dex in our age and metallicity estimates (with opposite sign) derived with scaled-solar models for an \afe~increase of 0.3~dex (see Sec.2.4.2 of G05).

We followed the same methodology as outlined in G05 for SDSS DR2 and for the SDSS DR4 parameters used in \cite{pasquali10}. However, SDSS DR7 stellar metallicities are on average higher by $0.03-0.1$~dex for masses below $\sim10^{10}M_\odot$ and lower by $0.03$~dex at larger masses with respect to the previous releases, for the galaxies in common, because of differences in \dn~measurements and associated uncertainties due to modifications in the spectro-photometric calibration. We note that these differences result in slightly shallower mass-metallicity relation and a slightly smaller offset in stellar metallicity between satellites and central galaxies with respect to what observed in \cite{pasquali10}, but do not affect our main conclusions.

In order to obtain an estimate of age and metallicity we require that all the five absorption indices are measured from the emission-line cleaned spectra. For a reliable estimate of these parameters, we further restrict the analysis to those galaxies with a spectral S/N of at least 20. The final sample of galaxies with a measure of the stellar population parameters and with $S/N>20$ comprises 113307 galaxies, of which 87284 are centrals and 26023 are satellites. The distributions in stellar mass and halo mass for the final galaxy samples are shown in Fig.~\ref{fig:mass_distr} (solid histograms) compared to the distributions for the original sample (dotted histograms). The cut in S/N affects central and satellite galaxies in a similar way, leaving 23\% and 25\% galaxies respectively of the corresponding original samples; it also preferentially excludes galaxies with lower stellar mass and, as a consequence, that reside in groups with lower halo mass.

\begin{figure}
\centerline{\includegraphics[width=9truecm]{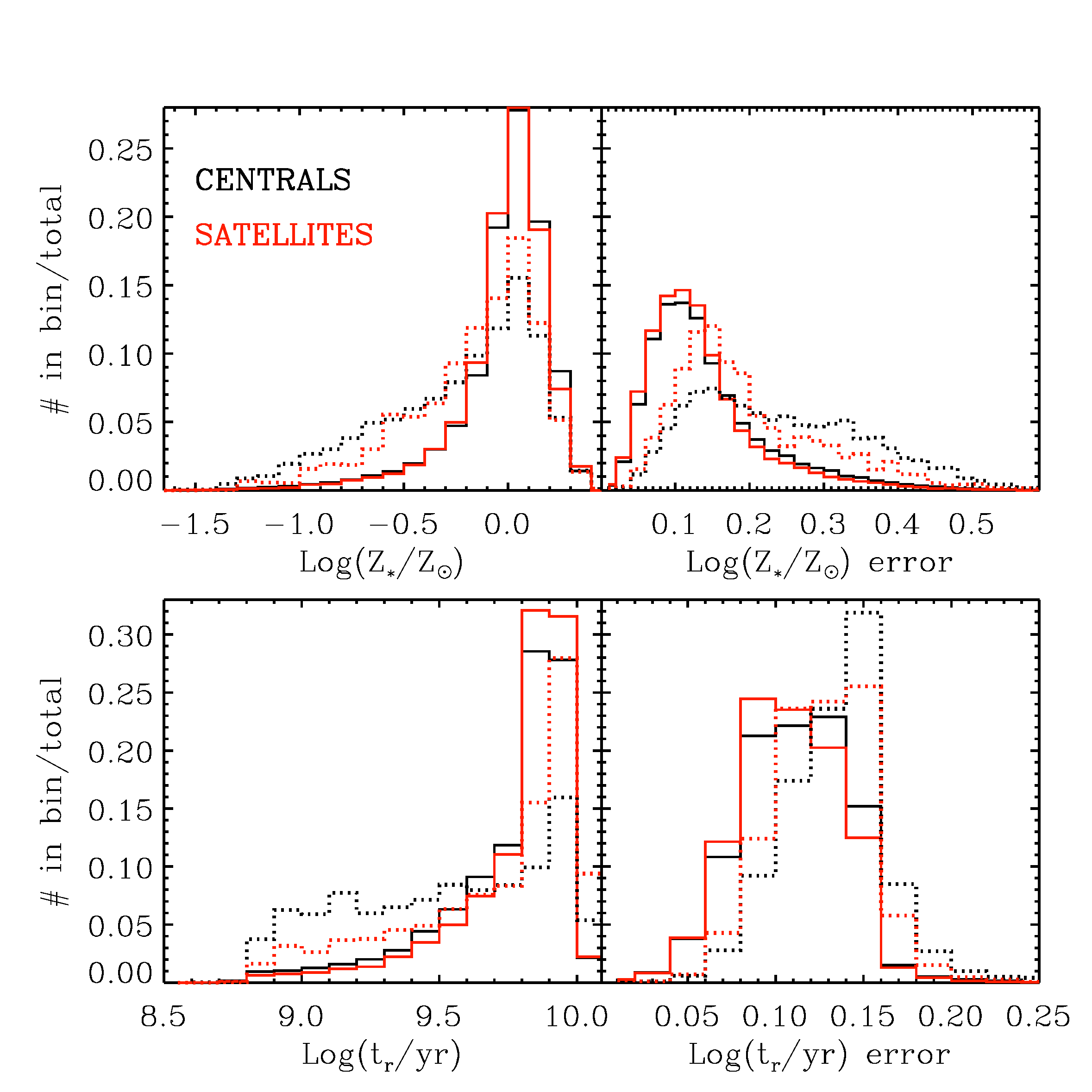}}
\caption{Distribution in derived stellar metallicity ({\it upper left panel}) and its uncertainty as given by half of the $84^{th}-16^{th}$ interpercentile range ({\it upper right panel}) for our high-S/N sample of `central' (black) and `satellite' (red) galaxies. The distributions in derived r-band light-weighted age and its associated uncertainty are shown in the lower panels for central and satellite galaxies. Each histogram is normalised to unity. The dotted histograms show the distributions obtained weighing galaxies by $1/V_{max}\times w_{SN}$.}\label{fig:distr_zstar_tstar}
\end{figure}

Figure~\ref{fig:distr_zstar_tstar} shows the distribution in the median-likelihood estimates of stellar metallicity and r-band light-weighted average age and their associated uncertainties, as given by half of the $84^{th}-16^{th}$ interpercentile range of their PDF, for the high-S/N samples of centrals (black) and satellites (red). The distributions in stellar metallicity and mean age for the two populations are very similar, although we notice that satellite galaxies are slightly more abundant at old ages than central galaxies. The median uncertainty on stellar metallicity is 0.12~dex, while that on light-weighted age is 0.11~dex for both samples. The dotted histograms show the distribution obtained by weighing galaxies in order to correct for Malmquist bias and for incompleteness due to the S/N cut (see Sec.~\ref{sec:rel_mstar}). This shows that galaxies with younger and/or metal-poorer populations are more affected by the cut in S/N.

\subsection{\afe~abundance ratios}\label{sec:afe}
We adopt a semi-empirical estimate of \afe~building on the approach of \cite{gallazzi06} (hereafter G06). For each galaxy we measure the index
ratio \mgtfe~on the observed spectrum\footnote{$\rm \langle Fe\rangle$ is the average of the Fe5270 and Fe5335 index strengths.} and the difference \Dmgfe~between this value and the index ratio measured on all the models in the library. We then build the PDF of \Dmgfe~fitting the \afe-independent features \dn, \hb, \hdg, \mgfep, \mgtwofe, as we do for age and metallicity. The excess \Dmgfe~in the data with respect to the BC03 base
models can be interpreted as an excess of \afe~with respect to solar. \footnote{We note that the BC03 SSP models follow the abundance pattern of the Milky Way, hence they are truly \afe$=0$~only for [Z/H]$\gtrsim0$. The derived abundance ratios should then be regarded as relative to the MW abundance pattern. We found that the zero-point of our \afe~estimates may be biased low for $\log(Z/Z_\odot)<-0.4$, but this does not affect our main results (see Appendix~\ref{A1}).}
We note that these measures are fully consistent with \Dmgfe~measured on the bestfit model only, as done
in G06, with a rms scatter of 0.05 comparable to the average $16th-84th$ interpercentile half-range of
the \Dmgfe~PDFs (0.025). The average uncertainty on the observed index ratio \mgtfe~is 0.14. We combine in quadrature the two contributions (width of PDF and observational error) to estimate the uncertainty on each \Dmgfe~measurement (see Appendix~\ref{A1}).

Considering the excess \Dmgfe, rather than the observed index ratio, significantly removes the dependence on age and metallicity.
Indeed, we have checked that the relation between \Dmgfe~and \afe~is largely independent of age and metallicity for a wide range in these
parameters (see also Sec. 2 of G06) using the SSP models of \cite{thomas03} and \cite{thomas04}, hereafter TMK04\footnote{These models have been made available by Daniel Thomas
at www.dsg.port.ac.uk/\~thomasd/tmb} (see Appendix~\ref{A1}). Specifically, we consider five metallicities ([Z/H]$=-1.35,-0.33,0.0,0.35,0.67$) and
seventeen age values between 0.6 and 15~Gyr. For each age and metallicity we compute the difference between the \mgtfe~index of models with
\afe$=-0.3,0.3,0.5$ with respect to the index strength of the solar-scaled model (\afe=0) with the same age and metallicity.
The rms scatter in \afe~for a given value of \Dmgfe~at fixed metallicity and for age varying between 1 and
15~Gyr is below 3\% for metallicities larger than $[Z/H]=-0.33$ and 8\% for $[Z/H]=-1.35$. Similarly, the rms scatter at fixed age for metallicity
varying between -1.35 and 0.67 is below 10\% for ages younger than 4~Gyr and around 13\% for older ages.

Contrary to G06 where only early-type galaxies were analysed, the sample used in this study includes both star-forming and quiescent galaxies and covers larger age and metallicity ranges, where the relation between \Dmgfe~ and \afe~mildly changes. Therefore, we explicitly translate the empirical \Dmgfe~diagnostic into an estimate of \afe. In order to do so, 
we calibrate the relation between \Dmgfe~and \afe~as a function of age and
metallicity adopting as default the TMK04 models. For each age and metallicity, we fit a third order polynomial
over the \Dmgfe~and \afe~range of the models. For each galaxy in the sample we apply the fitted function corresponding to the age and metallicity
closest to the estimated ones (see Sec.\ref{sec:age_met}). For those galaxies whose \Dmgfe~value lies outside the model range we adopt the linearly extrapolated function. This
happens for only 6\% of the sample and correspond to \afe$\gtrsim0.5$ and we note that the results of this work are not influenced by these
galaxies. We notice that the fitted \Dmgfe-\afe~relations have a very small dependence on age and metallicity in particular for ages $>1$Gyr or metallicities $\gtrsim0.5 Z_\odot$.\footnote{Systematic uncertainties on age/metallicity of $\sim0.05$~dex would translate into \afe~differences of $<0.01$~dex. This would be the case even for variations in age of a few Gyr.}
We have checked with other sets of models \citep{TMJ11,walcher09,Vazdekis15} that we find very similar relations between \Dmgfe~and \afe~and that \Dmgfe~has a negligible dependence on age and metallicity (see Appendix~\ref{A1}). The conclusions presented in the following are not affected by the chosen calibration: we find consistent scaling relations between \afe~and stellar mass and halo mass and, in particular, a consistent comparison between centrals and satellites split into halo or stellar mass bins.

\begin{figure}
\centerline{\includegraphics[width=9truecm]{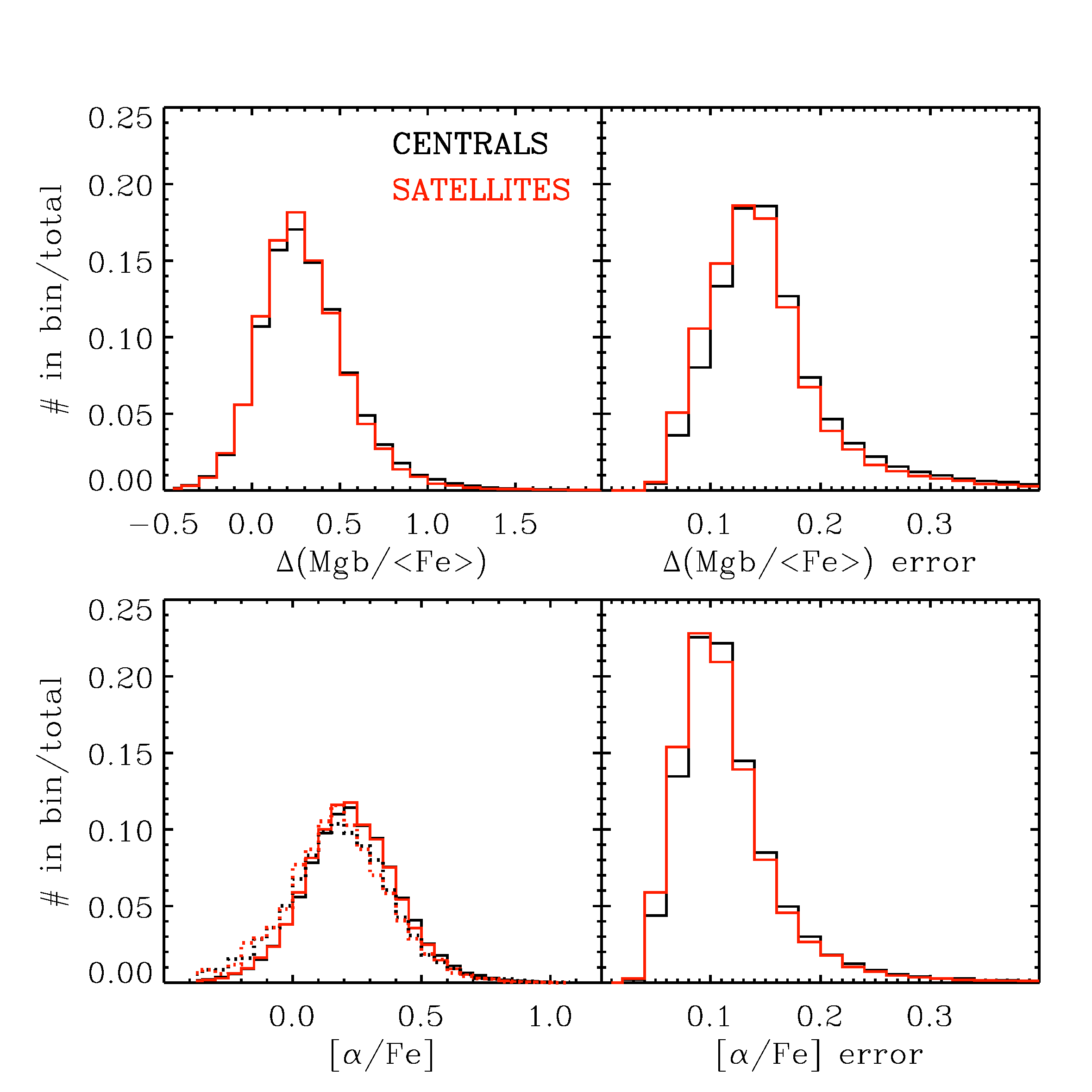}}
\caption{Distribution in the empirical estimator \Dmgfe~ ({\it upper left panel}) and in the inferred \afe~({\it lower left panel}) for satellites (red histograms) and for central galaxies (black histograms). The right-hand panels show the distributions in the associated uncertainties combining the observational error on \mgtfe~index and the $84^{th}-16^{th}$ interpercentile half-range of the PDF of \Dmgfe~or \afe. Each histogram is normalised to unity. The dotted histograms in the bottom-left panel show the distributions obtained weighing galaxies by $1/V_{max}\times w_{SN}$.}\label{fig:distr_alphafe}
\end{figure}

Figure~\ref{fig:distr_alphafe} shows the distributions in the empirical estimator of element abundance ratio \Dmgfe (upper panel) and in the derived \afe~estimates (lower panel) for central galaxies (black) and satellites (red). The distributions for the two galaxy samples are remarkably similar, despite the small offsets observed in the mass distributions. The right-hand panels show the uncertainties on \Dmgfe~and \afe, combining the observational uncertainty on the \mgtfe~index strength and the  $84^{th}-16^{th}$ interpercentile half-range of the PDF. 

\subsection{Statistical corrections}
The original galaxy sample is not volume limited and therefore suffers from Malmquist bias. In order to correct
for this bias, we weigh each galaxy by the inverse of the maximum visibility volume (V$_{max}$), i.e. the
comoving volume out to a comoving distance at which the galaxy would pass the magnitude limit of the survey. In
addition to this, the cut in S/N that we apply for reliable estimates of stellar population parameters introduces
a further bias toward brighter galaxies at fixed mass. We wish to correct for incompleteness 
as a function of stellar population properties at fixed stellar mass, so as to properly weight galaxies in the scaling relations. 
In order to do so we estimate the S/N selection bias as a function of both stellar mass and absolute $g-r$ color, 
taken as proxy of galaxy physical parameters. Specifically, we compute the number ratio between the original sample and
the final high-S/N sample in bins of stellar mass and absolute $g-r$ Petrosian color of width 0.1~dex and 0.2~mag respectively, for satellites and central galaxies separately.  We
thus weight each satellite and central galaxy for this ratio, $w_{SN}$, to correct for the galaxies missed due to the S/N cut. We do not consider bins in which high-S/N galaxies are 
fewer than 10 and are less than 10\% of the number of galaxies before the S/N cut, and we put the weight to zero for galaxies 
falling in those bins. They are in any case a negligible fraction of both the initial and final sample. 
Unless otherwise stated, in all the relations shown in this work the galaxies are weighted by $1/V_{max}\times w_{SN}$. We have
checked that our results do not critically depend on these weights, in particular the differences between central
and satellite galaxies are qualitatively the same if no weight is applied.

Another limitation of the spectroscopic data is that they are taken through fixed-aperture fibers and hence sample different spatial extent of the galaxies. Quantifying this effect on a galaxy-by-galaxy basis is uncertain and depends on several factors, and we do not attempt to correct for it. However, we have checked that the trends discussed in the next sections (in particular Sec.~\ref{sec:rel_SFR}) are not driven by aperture bias due to galaxies with higher mass and in higher mass halos extending to higher redshift (Appendix~\ref{A2}).
 
\section{Trends with stellar mass}\label{sec:rel_mstar}
\begin{figure}
\centerline{\includegraphics[width=9truecm]{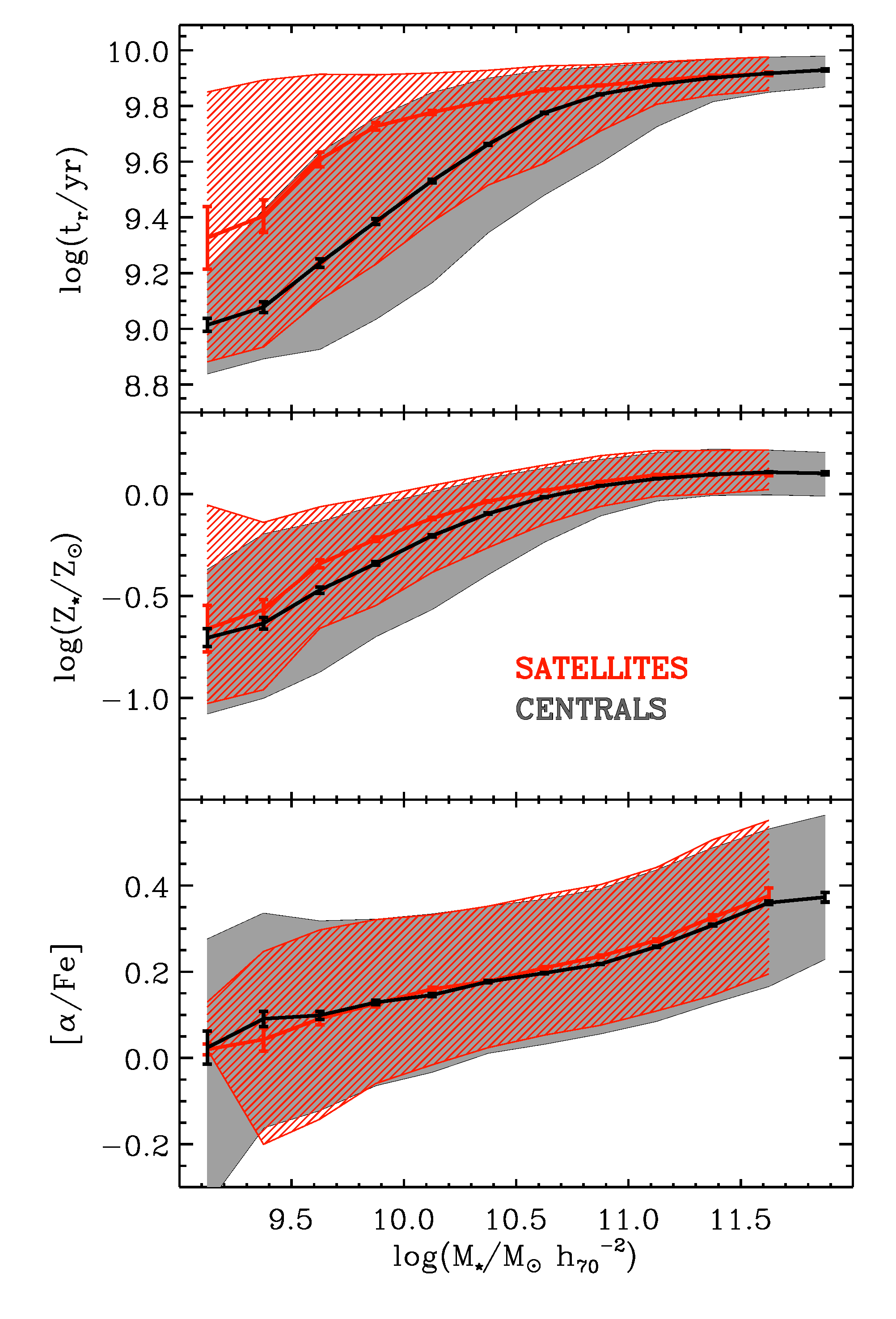}}
\caption{Distribution in r-band light-weighted age (upper panel), stellar metallicity (middle panel)
and \afe~(bottom panel) as a function of stellar mass for central (black curve and grey region) and satellite
galaxies (red curve and hatched region). The solid lines represent the median trend, with {\it error bars indicating the
error on the median in each bin of stellar mass}. The hatched regions represent the
$16^{th}-84^{th}$ interpercentile range of the distributions. Only galaxies with spectral $S/N\geq20$
are included. Each galaxy is weighted to correct for Malmquist bias and for incompleteness due to the S/N cut.}\label{fig:rel_mstar}
\end{figure}

It is known that both the mean stellar age and the stellar metallicity in galaxies increase on average with galaxy mass \citep[e.g.][]{gallazzi05,Thomas05,mateus06,Rosa15}. The element abundance ratio \afe~is also known to correlate with galaxy mass for early-type/quiescent galaxies \citep[e.g.][]{jorgensen99,trager2000,Thomas05,gallazzi06,graves09b,Conroy14,walcher15}. In this section we analyse the scaling relations between light-weighted age, stellar metallicity and, for the first time, \afe~for all galaxy types versus galaxy stellar mass. We explore whether these scaling relations depend on galaxy hierarchy. In Fig.\ref{fig:rel_mstar} we
plot the median trends (solid line with error bars) of age, stellar metallicity and \afe~as a function of stellar
mass for satellites (red) and central galaxies (black). The error bars indicate the error on the median\footnote{Because of the non-gaussianity of the distributions and the weights applied, instead of using the standard definition of the error on the median, we compute this as $1.25\times(P84-P16)/2/\sqrt{N}$, where N is the number of galaxies in each stellar mass bin.}, while the
shaded regions indicate the $16^{th}-84^{th}$ interpercentile range of the distributions. 

From the upper panel of Fig.\ref{fig:rel_mstar} it is immediately clear that at masses lower than $\rm 10^{10.8}M_\odot$ satellite galaxies are older than equally massive central galaxies. The difference in age between satellites and central galaxies increases with decreasing stellar mass at least down to $\rm \sim10^{9.4}M_\odot$ reaching $0.35-0.4$~dex. A similar but weaker trend is observed for stellar metallicity (middle panel), such that satellites are metal-richer than centrals by $\sim0.15$~dex in the mass range $\rm 10^{9.6}-10^{10.6}M_\odot$. Over this mass range, these differences in metallicity are significant between 5 and 13$\sigma$~level compared with the error on the median, but the significance drops at masses lower than $\rm 10^{9.5}M_\odot$. At masses above $\rm 10^{10.5}M_\odot$ we find a tiny difference in the median age and the median stellar metallicity of satellite and central galaxies, but still statistically significant up to masses of $10^{11.12}M_\odot$, above which centrals and satellites are indistinguishable as already shown in \cite{pasquali10}. Not only the median trends are different for satellites and for centrals, but also the lower and upper percentiles. In particular we notice a lower fraction of young or metal-poor galaxies among satellites, and a larger spread in age toward old ages for low-mass satellites with respect to centrals. These results reproduce those already shown in \cite{pasquali10} but here updated with the SDSS DR7 measurements.

In addition to age and metallicity, in this work we explore the dependence of \afe~on stellar mass and galaxy hierarchy for 
all galaxy types. The bottom panel of Fig.\ref{fig:rel_mstar} shows that \afe~increases with stellar mass also when the whole galaxy population is considered and not only for early-type galaxies. Specifically we find that \afe~increases from $\sim0.05$~dex at $\rm 10^{9.1}M_\odot$ to $\sim0.35$~dex at $\rm 10^{11.5}M_\odot$. Contrary to age and metallicity, both the median trend and the 16-84 percentile range of \afe~are virtually indistinguishable for satellites and central galaxies. We notice, though, that in the mass range $\rm 10^{10.6}-10^{11.4}M_\odot$~the tiny excess in \afe~of satellites is statistically significant at the $4-7\sigma$~level (compared with the error on the median). This will be better explored in Sec.~\ref{sec:rel_SFR}.

The differences in light-weighted mean age suggest at face value that satellite-specific processes have influenced and suppressed the more recent star formation history of relatively low-mass satellites to a larger extent than what has occurred for their central counterparts. These mechanisms have also led, on average, to a higher degree of metal enrichment of satellites' stellar populations. However, the element abundance ratio \afe~is not affected suggesting that quenching happens (or terminates) late enough for Fe-peak elements to be incorporated into stars. To further elucidate the possible origin of these difference, we investigate in the next sections how they depend on halo mass and on galaxy type.

%\clearpage

\section{Trends with halo mass}\label{sec:rel_mhalo}
In Fig.\ref{fig:rel_mstar_mhalo} we explore any further dependence of stellar population parameters on the group halo mass. 
The left-hand (right-hand) panels show the relations with stellar mass (halo mass) for galaxies residing in halos of different mass (in bins of $M_\ast$). The median trends as a function of stellar mass (halo mass) of satellite galaxies in bins of halo mass (stellar mass) are shown by the solid coloured lines, with error bars indicating the error on the median. The median trends as a function of stellar mass (halo mass) of central galaxies in bins of halo mass (stellar mass) are shown by the dot-dashed coloured lines. The grey region and the white solid line in each panel show the median and percentiles of the overall distribution for centrals.

\subsection{Central galaxies}\label{sec:cen_mhalo}
For central galaxies, both age and stellar metallicity depend strongly on stellar mass: both parameters increase with stellar mass over roughly two orders of magnitude, with a flattening only at masses $\gtrsim10^{11}M_\odot$ (left-hand panels). Centrals' age and metallicity are instead independent of halo mass except for halos less massive than $10^{12}M_\odot$ (right-hand panels). We note though that, according to the relation between host halo mass and central stellar mass \citep{yang08}, the halo mass range over which the relations are flat ($\rm M_h>10^{12.5}M_\odot$) roughly corresponds to the range over which also the relations with stellar mass are flat ($\rm M_\ast>10^{11}M_\odot$). On the contrary, we do not detect any significant flattening in the scaling relations of \afe~of central galaxies: their \afe~shows a similar monotonic increase with both stellar mass and halo mass (bottom panels). We note that the slope of the relations with stellar mass (halo mass) for central galaxies is independent of halo mass (stellar mass): centrals in different halos (dot-dashed colored lines) follow the same trends with stellar mass as the global central galaxies populations. We only note a milder increase of age and metallicity with halo mass at fixed stellar mass for $\rm M_h<10^{12.5}M_\odot$ with respect to the global trends, but the range in halo mass probed, at fixed M$_\ast$, is narrow.

\cite{Trussler20b} discuss a feature of central galaxies at fixed stellar mass being slightly more metal-poor in higher mass halos. We do not find such a feature. Instead, over the small mass range of overlap, we see that at fixed stellar mass centrals in more massive halos tend to be either equal or slightly more metal-rich, older and more $\alpha$-enhanced than those in lower-mass halos: note the positive slope with M$_h$ of the dot-dashed lines in the right panel of Fig.~\ref{fig:rel_mstar_mhalo} and the systematic vertical offsets of the lines at fixed halo mass in the left panels (these trends are more clearly visible in Fig.\ref{fig:rel_cen_zoom} zooming onto the high-mass regime of centrals). It is a subtle effect and the difference could come from the different physical parameter estimates used and/or from the different binning in halo mass.

The slight increase of metallicity with M$_h$ for central galaxies is consistent with that found, at fixed velocity dispersion (rather than M$_\ast$), by \cite{labarbera14} for the population of early-type galaxies only. Notice that, instead, the trends of age and \afe~with M$_h$ seem to differ with respect to those of \cite{labarbera14}, who found that, at fixed velocity dispersion, early-type centrals in massive groups are younger and less $\alpha$-enhanced than early-type centrals in low-mass halos. These differences may be due to the fact that our trends include the contribution of both early- and late-type galaxies, the latter dominating the low-mass range, while \cite{labarbera14} considered only (morphological and color-selected) early-type galaxies. Moreover, \cite{labarbera14} compared galaxy properties as a function of velocity dispersion rather than stellar mass as in the present work.

\subsection{Satellite galaxies}\label{sec:sat_mhalo}
Concerning satellite galaxies, all the stellar population parameters studied here increase with galaxy stellar mass even at fixed halo mass, suggesting that stellar mass is the primary driver of stellar population age and chemical abundances (left-hand panels of Fig.~\ref{fig:rel_mstar_mhalo}). This is not new and evidences of this statement have been reported by several works \citep[e.g. G06,][]{pasquali10,peng10,Thomas10,Trussler20b}. However, in addition to stellar mass dependence, it is evident that the light-weighted mean age also strongly depends on halo mass at fixed stellar mass: for satellites less massive than $\sim10^{11}M_\odot$ their mean stellar age increases on average with increasing halo mass and the relation becomes steeper 
for lower stellar masses (upper-right panel). Low-mass satellites residing in more and more massive halos deviate to older and older ages with respect to equally-massive central galaxies, such that below $10^{10}M_\odot$ satellites in $\rm M_h>10^{14}M_\odot$ are $\sim0.6$~dex older than satellites in $\rm M_h<10^{12.5}M_\odot$ and than equally-massive centrals (upper-left panel).
The stellar metallicity of satellite galaxies depends primarily on galaxy stellar mass, but a mild dependence on halo mass at fixed stellar mass is detected for satellites less massive than $\sim10^{10.5}M_\odot$: the stellar metallicity of satellites increases by about 0.2~dex over a range in halo mass from $\sim10^{12}$ to $10^{14.5}M_\odot$ and with respect to equally-massive central galaxies (middle-right panel). No significant dependence on halo mass is observed for the \afe~of satellite galaxies (bottom panels): the \afe$-M_\ast$ relation is the same for satellites in different environments, except for a small \afe~excess for satellites residing in the most massive halos ($\rm M_h>10^{14}M_\odot$).

These trends in stellar age suggest that for low-mass galaxies the mass of the halo onto which they accrete influences the process of star formation quenching, being it facilitated or occurring earlier in more massive halos. The excess in stellar metallicity of satellite galaxies can be explained in a scenario in which the quenching is associated with the suppression of metal-poor gas inflows in high density environments, as a result of processes that remove the gas in the galaxy halo and/or in the galaxy disc, as discussed in \cite{bahe17} and \cite{Trussler20a}. If we interpret \afe~as a proxy for the timescale of chemical enrichment, its  lack of environmental dependence may indicate that any environmental quenching should take long enough or occur after long enough time to allow the reprocessing of Fe-peak elements. This interpretation is in line with the long timescale of SF suppression inferred in \cite{pasquali19} from the gradients in specific SFR versus phase-space zone. 

{\it Overall, the observed global trends in age, the small stellar metallicity excess for low-mass satellites in massive halos and the lack of significant \afe~differences as a function of group hierarchy and halo mass, manifest the action of environment on the more recent star formation and not the bulk star formation in the past.}

\subsection{Comparing centrals and satellites at fixed stellar and halo mass}\label{sec:cen_sat_mhalo}
By fixing both the stellar mass and the halo mass, there are no significant systematic differences between centrals and satellites. The similarity of centrals and satellites at fixed stellar and halo mass has already been discussed in \cite{wang18} who interpret it as due to a similarity in the quenching process affecting centrals and satellites in their host halo. However, because of the relation between group halo mass and $M_\ast$~for centrals \citep{yang08}, central galaxies span a narrow range in stellar mass at fixed $\rm M_h$ and there is little overlap with satellites (except for $M_h \lesssim 10^{12.5}M_\odot$). Conversely, except for $\log M_\ast/M_\odot=10.5-11.5$, at fixed stellar mass there is a limited range of overlap in $\rm M_h$ between centrals and satellites. Controlling for both stellar and halo mass, satellites and centrals appear to follow similar scaling relations, but covering different mass ranges (comparing solid and dot-dashed colored lines). This has been interpreted by \cite{wang18} as an indication that satellite-specific environmental effects ('nurture') are not important. However this does not fully explain the different distributions of centrals and satellites that we observe. Moreover, for $\log M_h/M_\odot<12.5$ (the only environment where satellites and centrals span a similar stellar mass range) satellites are $\sim0.05$~dex older than centrals at fixed stellar mass (black solid versus dot-dashed curves in the left-hand panels). For $\log M_\ast/M_\odot=10-11$, instead, centrals are $\lesssim0.05$~dex older and more metal-enhanced than satellites at fixed halo mass (green and blue dot-dashed versus solid curves in the right-hand panels). However this could be due to the larger average $\rm M_\ast$ of centrals in each $\rm M_h$ bin and the relation between stellar population parameters and stellar mass. Indeed, we checked that for each stellar mass bin and as a function of $\rm M_h$ centrals are on average $0.2$~dex more massive than satellites. 

\begin{figure*}
\centerline{\includegraphics[width=15truecm]{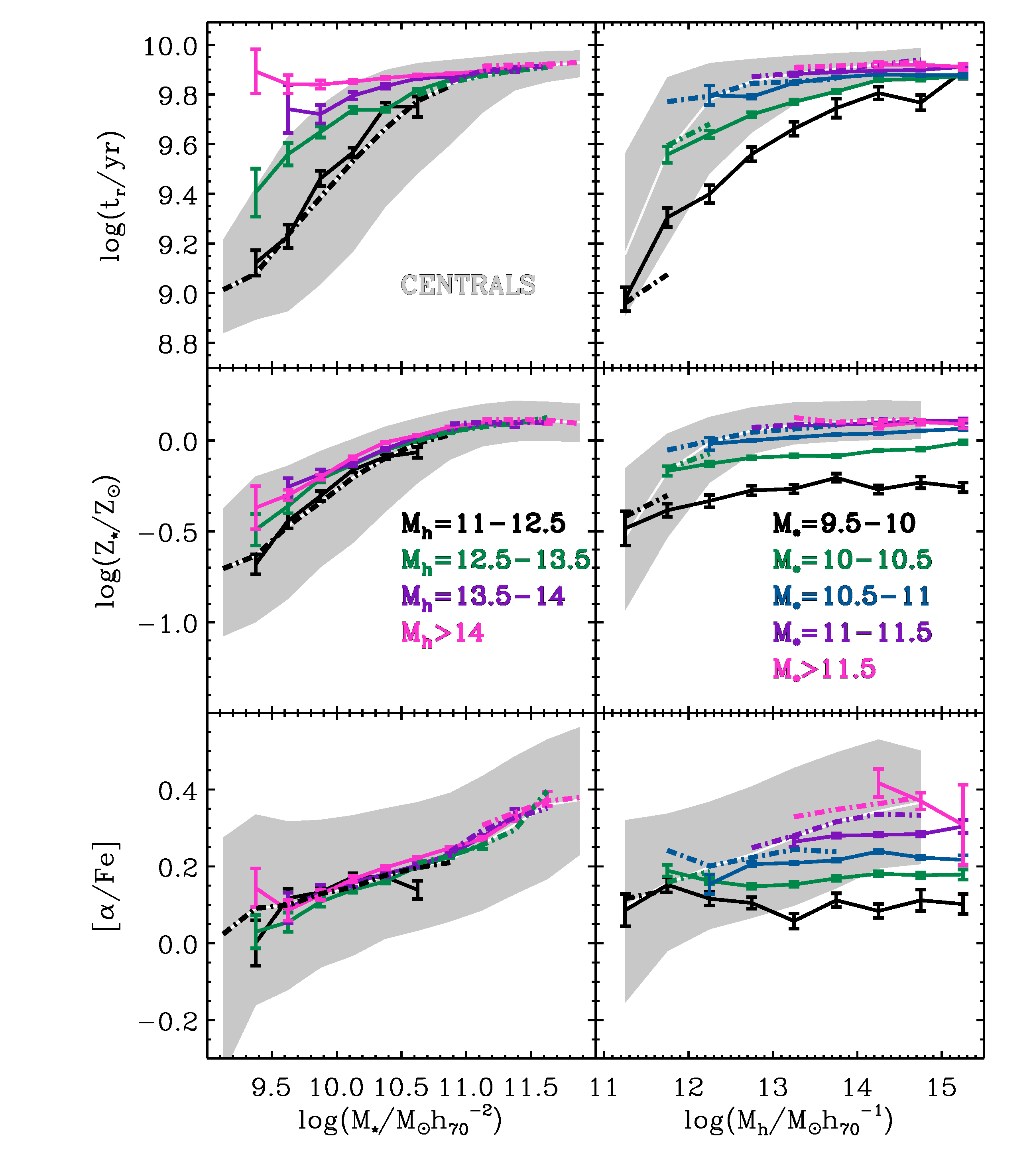}}
\caption{{\it Left:} Distribution in light-weighted age, stellar metallicity and \afe~as a function of 
stellar mass for central galaxies (grey shaded region enclosing the 16-84 interpercentile range) and for satellite galaxies in different bins of 
log halo mass (colored lines). Centrals are also divided in the same halo mass bins and the corresponding trends are indicated by the dot-dashed lines
 {\it Right:} Distribution in light-weighted age, stellar metallicity and \afe~as a function 
of halo mass for central galaxies (grey shaded region enclosing the 16-84 interpercentile range) and for satellite galaxies in different bins of log 
stellar mass (colored lines). The dot-dashed lines show the trends for centrals in the same stellar mass bins. Only bins with at least 20 galaxies are shown. 
The error bars associated with the trends of satellite galaxies are the {\it error on the median}.}
\label{fig:rel_mstar_mhalo}
\end{figure*}

We should keep in mind that when fixing both M$_h$ and M$_\ast$ we are considering satellites that have a similar M$_\ast$ to their own central as well. In this regime misclassifications of centrals due to stellar mass errors and to the more physical fact that the most massive galaxy is not always the one sitting at the bottom of the potential well \citep{skibba11} can be important and wash out true differences between centrals and satellites.
{\it What emerges from Fig.~\ref{fig:rel_mstar_mhalo} is that galaxies in the local Universe in order to be old and chemically evolved need to be either very massive or be satellites in massive haloes.}

\section{Dependence on specific SFR}\label{sec:rel_SFR}
So far we have compared satellites and central galaxies irrespective of their current star formation activity. The older average 
stellar age of satellites with respect to equally-massive central galaxies can be to first order the result of a higher fraction of 
passive galaxies among the satellite population. Indeed the frequency of passive galaxies is a strong function not only of galaxy mass 
but also of environment \citep[e.g.][]{peng12}. To quantify this effect in our data we classify the galaxies in our sample on the basis of their specific 
SFR. 

\begin{figure}
\centerline{\includegraphics[width=9truecm]{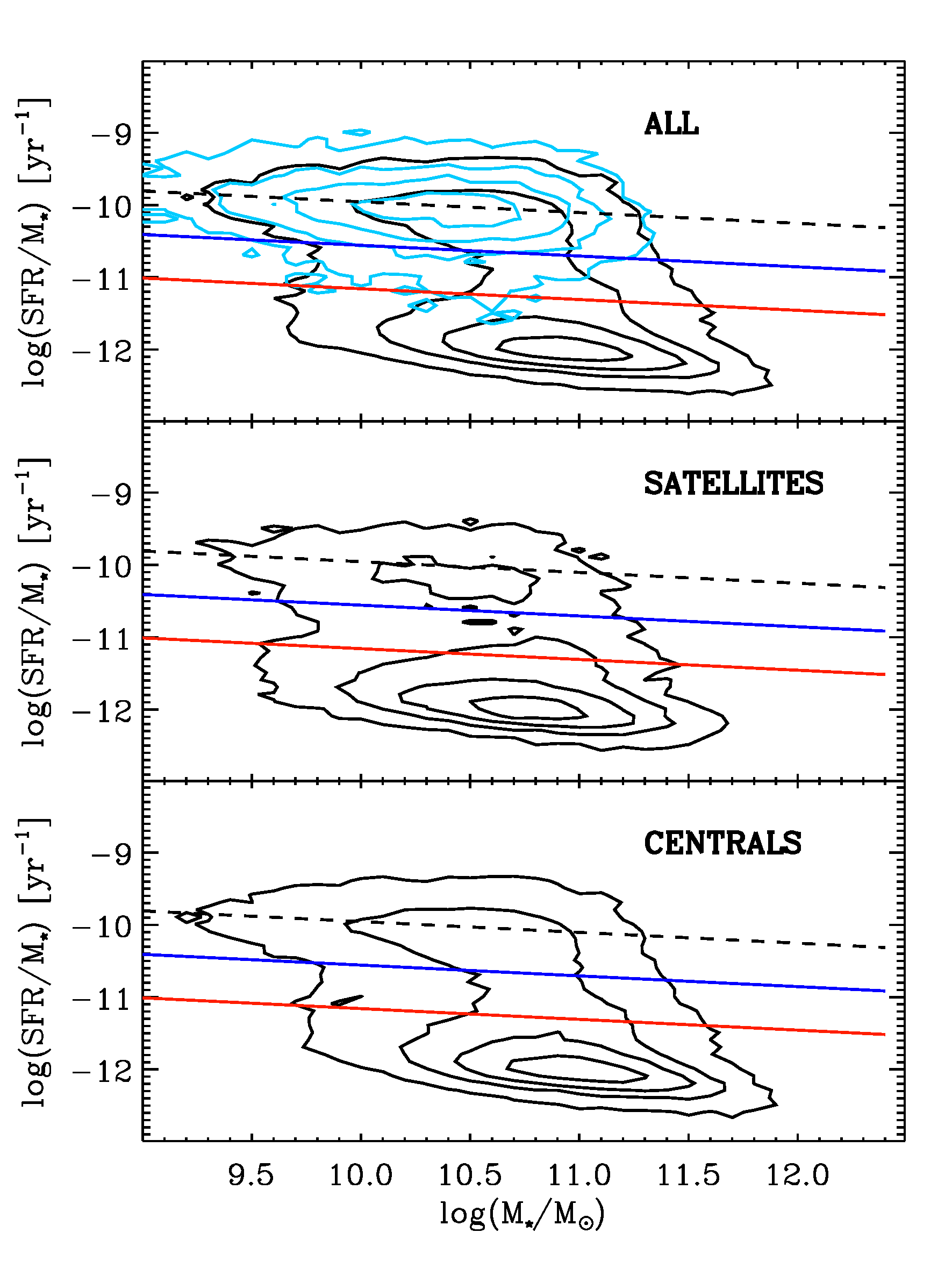}}
\caption{Distribution in specific SFR as a function of stellar mass for the whole sample (upper panel), for 
satellites (middle panel) and central galaxies (lower panel). In each panel the dashed line shows the linear relation fit for all galaxies (independent of 
hierarchy) classified as star-forming based on the BPT diagram (cyan contours), the red and blue lines show the same relation offset by 2 and
4 $\sigma$~respectively. We define as quiescent those galaxies below the red line, as star-forming those above the blue line and as green-valley those in between.}\label{fig:ssfr_mstar}
\end{figure}

\begin{figure}
\centerline{\includegraphics[width=9truecm]{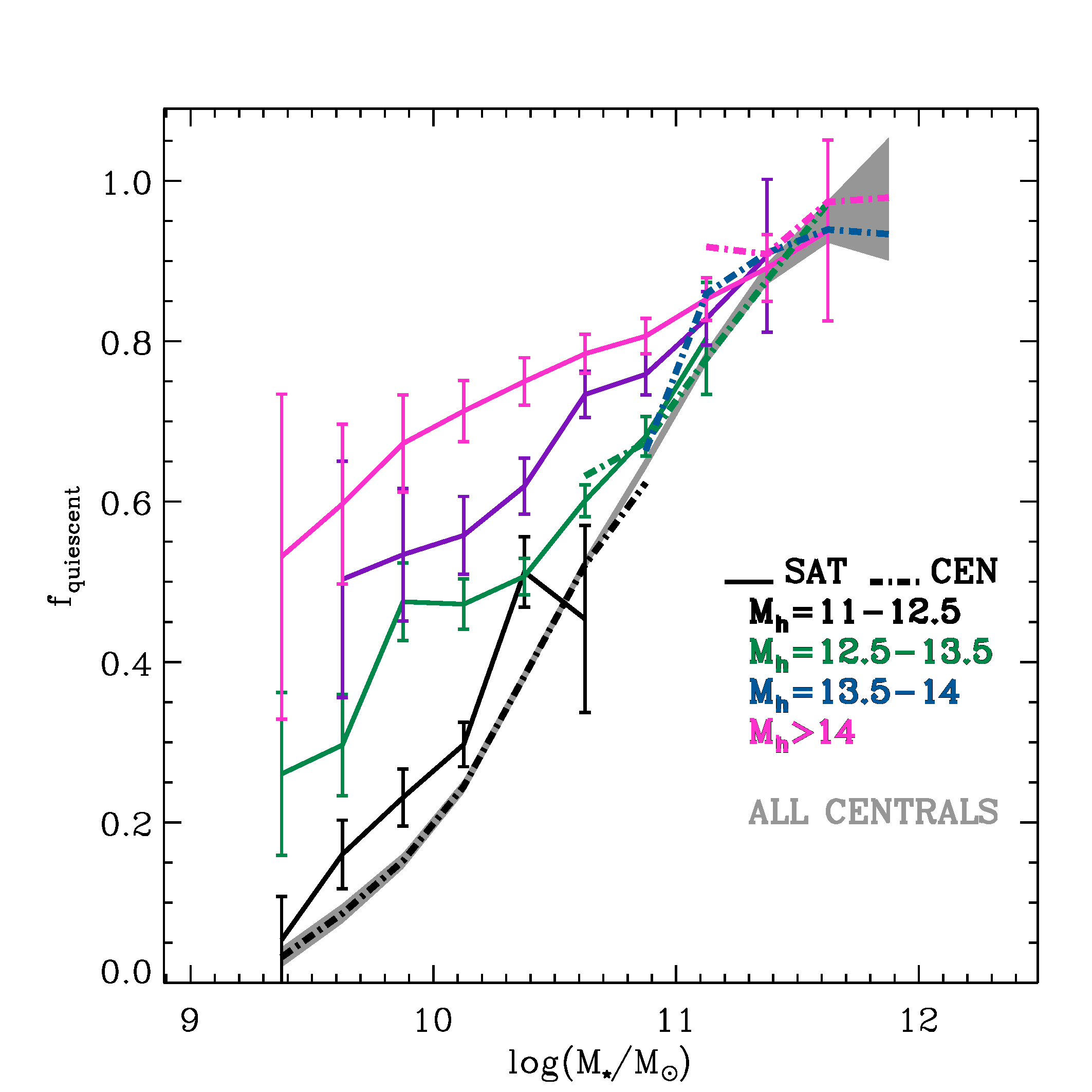}}
\caption{Fraction of quiescent galaxies as a function of stellar mass. Galaxies are weighted by $\rm1/V_{max}\times w_{SN}$. The grey shaded region refers to
central galaxies, while colored solid (dot-dashed) lines refer to satellite (central) galaxies binned according to their host
halo mass. Error bars indicate the Poissonian uncertainty. }\label{fig:fracQ}
\end{figure}

Figure~\ref{fig:ssfr_mstar} shows the distribution in specific SFR versus stellar mass for the whole sample (upper panel), and for satellites and central galaxies separately (middle and lower panels). The SFR values used here are those estimated for SDSS DR7 as in 
\cite{Jarle04} and aperture corrected to total values\footnote{We downloaded the catalogs from http://wwwmpa.mpa-garching.mpg.de/SDSS/DR7/sfrs.html where also a description of the method can be found.}. The dashed line in Fig.~\ref{fig:ssfr_mstar} shows the linear relation fit between specific SFR and stellar mass 
for galaxies in the whole sample which are classified as star-forming based on the \cite{BPT} (BPT) diagram (cyan contours). We define as quiescent those galaxies whose specific SFR is below this relation by more than $4\sigma$ (i.e. below the red line), star-forming those whose specific SFR extends from less than $2\sigma$ below the relation upward (i.e. above the blue line), and green-valley those with a specific SFR in between these values.

In Fig.~\ref{fig:fracQ} we show how the fraction of quiescent galaxies varies as a function of stellar mass for central galaxies (gray region) and for centrals and satellites divided into bins of halo mass (dot-dashed and solid colored lines, respectively). The fraction of quiescent galaxies increases to $>90\%$ at stellar masses above 
$10^{11.5}M_\odot$. Furthermore, at any given stellar mass down to $10^{9.5}M_\odot$ the quiescent fraction is higher among satellite galaxies and it increases with host halo mass, such that in halos more massive than $10^{14}M_\odot$ more than 50\% of satellites are quiescent at any fixed stellar mass. This is in agreement with what shown by many other works, irrespective of the exact definition of `quiescence' and of `environment' \citep[e.g.][]{baldry06,peng12,wetzel13,Trussler20b, delucia19}. In analogy with these results, considering early-type galaxies only, \cite{Thomas10} found that the fraction of rejuvenated galaxies instead decreases with increasing mass and with increasing environmental density. For central galaxies there is no clear dependence on halo mass, but this can only be probed for a narrow range in halo mass at fixed stellar mass. At high stellar masses ($\rm >10^{11}M_\odot$) the fraction of quiescent galaxies is the same for centrals and for satellites, but at lower stellar masses satellite galaxies have a higher passive fraction than equally-massive centrals in equally-massive halos: this is in particular evident for centrals and satellites in halos less massive than $\rm 10^{12.5}M_\odot$.

\begin{figure*}
\centerline{\includegraphics[width=15truecm]{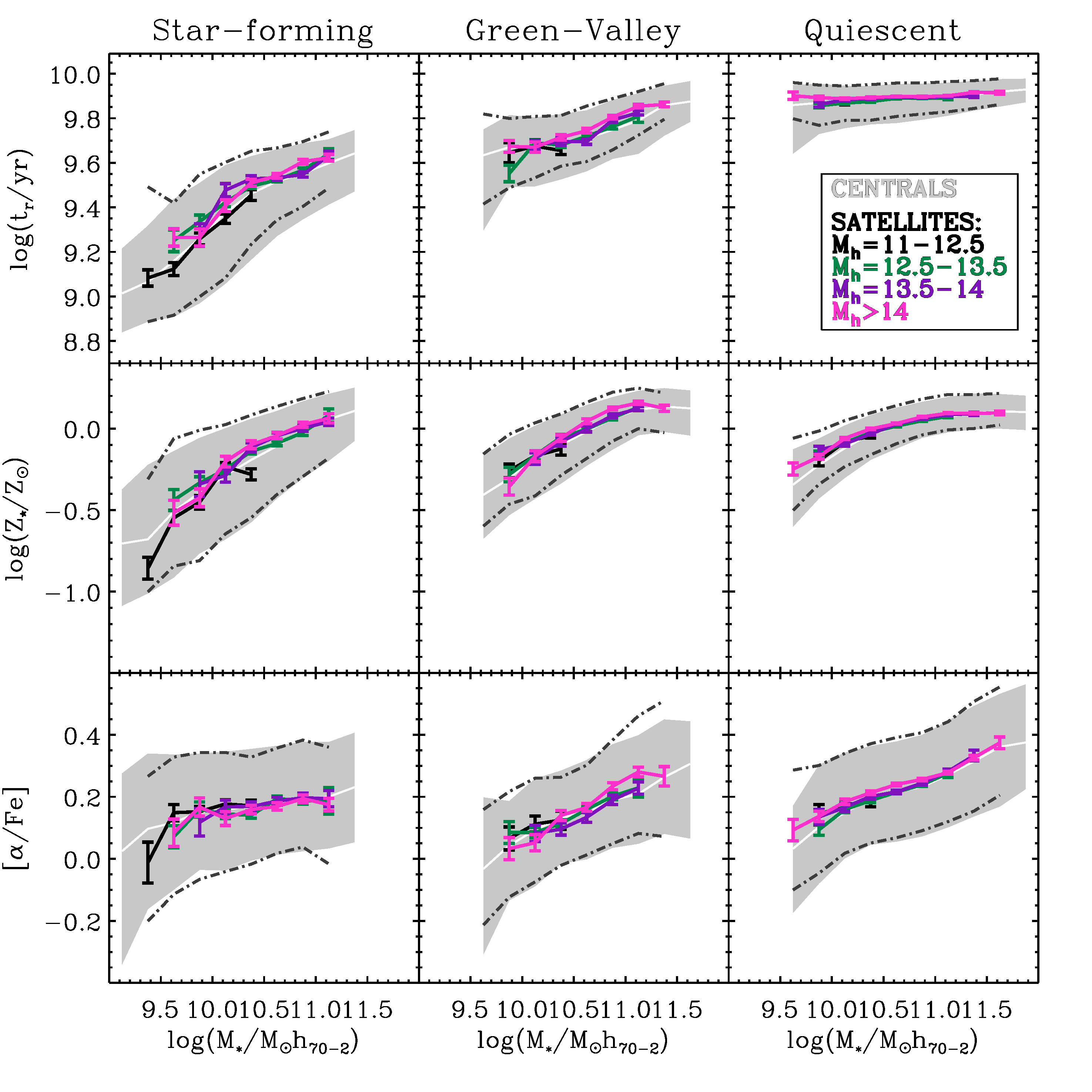}}
\caption{Relations between light-weighted age (upper panels), stellar metallicity (middle panels), \afe~(lower panels) and 
stellar mass for central galaxies (white line for the median and grey shaded region for the $\rm84^{th}-16^{th}$ percentile range) and for satellite galaxies split into bins of halo mass (colored lines with error bars giving the {\it error on the median}). Grey dot-dashed lines enclose the $\rm84^{th}-16^{th}$ percentile range of satellite galaxies without distinction on halo mass. Galaxies are divided into star-forming (left column), green-valley (central column) and quiescent (right column) according to their specific SFR. Only bins with at least 20 galaxies are shown.}
\label{fig:rel_mstar_sfr}
\end{figure*}

The increase in quiescent fraction with halo mass at fixed stellar mass qualitatively resembles the increase in average stellar age of satellite galaxies. The question arises whether the observed differences in age and metallicity between satellites in different halos and central galaxies are entirely due to the varying quiescent fraction or whether quiescent and star-forming satellites separately have different stellar population properties than their central counterparts owing to different star formation histories. We address this question in Fig.~\ref{fig:rel_mstar_sfr}, in which we show the relation between stellar age, stellar metallicity, \afe~and stellar mass for centrals (white line and grey region) and for satellites in different halo mass bins (colored lines) split into star-forming (left panels), green-valley (middle panels) and quiescent (right panels). The dot-dashed grey lines enclose the $84^{th}-16^{th}$~percentile range of all satellites. All the three stellar population parameters show a different dependence on stellar mass according to the galaxy star formation activity. Age and stellar metallicity increase steeply with stellar mass for star-forming galaxies, while their scaling relations become flatter for green-valley and quiescent galaxies. On the contrary, the \afe$-M_\ast$ relation is relatively flat for star-forming galaxies while it is significantly steeper for green-valley and quiescent galaxies.

In terms of environment, the first notable result from Fig.~\ref{fig:rel_mstar_sfr} is that when comparing galaxies with similar current star formation activity the distributions in stellar population properties at fixed stellar mass are very similar and in most cases virtually the same for satellite and central galaxies. {\it The differences observed in the scaling relations of satellites and centrals originate to a large extent from the dependence of quiescent fraction on stellar and halo mass}. Recently, \cite{Trussler20b} obtained the same results on the similarity in the age and stellar metallicity scaling relations of centrals and satellites when accounting for their quiescent fractions, using independent estimates of (mass-weighted) age and metallicity.

\begin{figure*}
\centerline{\includegraphics[width=15truecm]{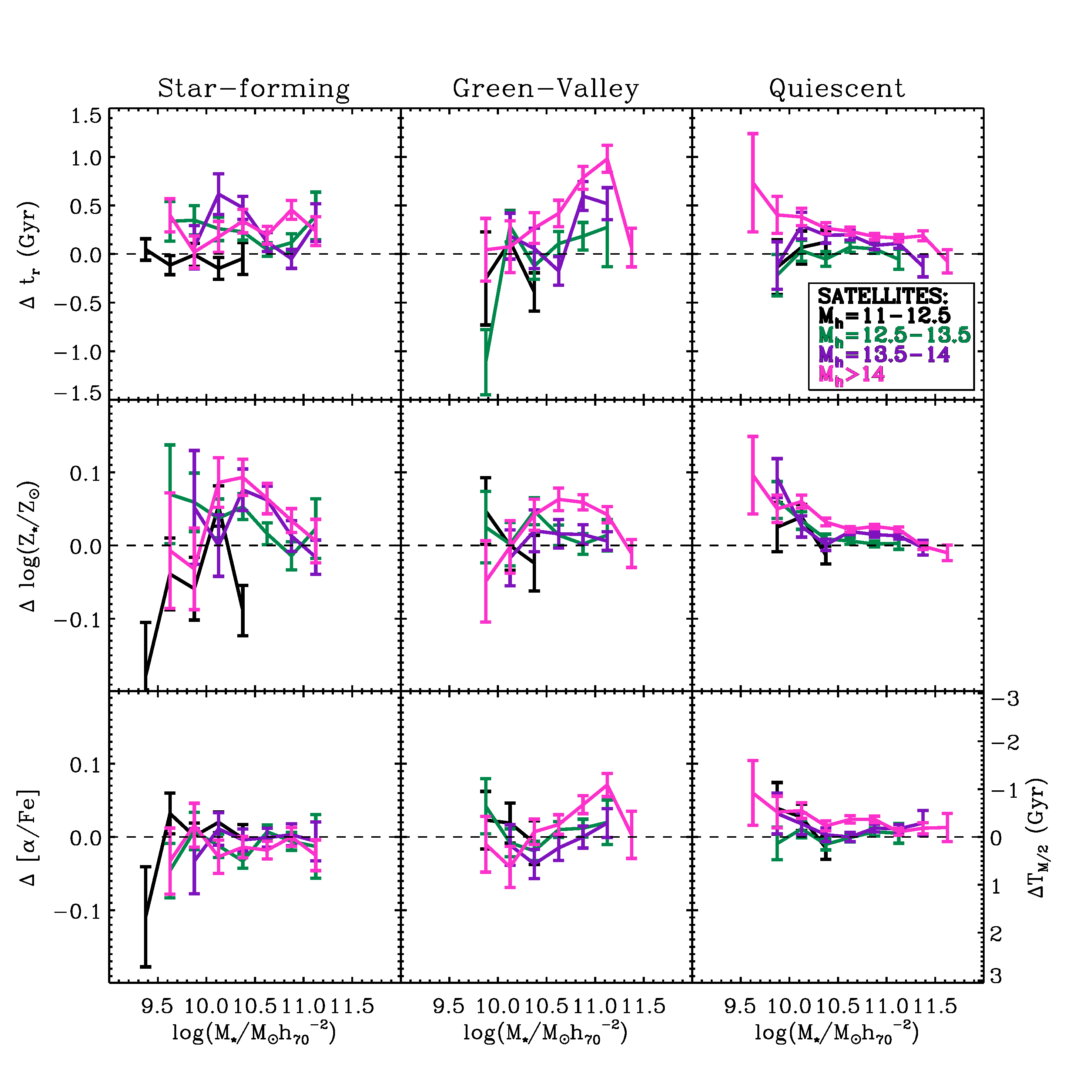}}
\caption{Difference in light-weighted age (upper panels), stellar metallicity (middle panels), \afe~(lower panels) between satellites in each halo mass bin (color coding as in Fig.\ref{fig:rel_mstar_sfr}) and central galaxies at fixed stellar mass. The errorbars show the quadrature sum of the errors on the median trends for satellites and centrals. Galaxies are divided into star-forming (left column), green-valley (central column) and quiescent (right column) according to their specific SFR. Only bins with at least 20 galaxies are shown. The y-axis label in the bottom right-hand plot indicates the half-mass time difference corresponding to a given \afe~difference according to the relation of de la Rosa et al (2011).}
\label{fig:resid_mstar_sfr}
\end{figure*}

We further take a closer inspection to the differences (or lack thereof) between satellites and central galaxies with similar star formation activity by exploring any dependence on halo mass. We divide satellite galaxies into bins of halo mass: the median relations with stellar mass are shown by the colored lines in Fig.~\ref{fig:rel_mstar_sfr}, while Fig.~\ref{fig:resid_mstar_sfr} shows the difference in age, metallicity and \afe~of satellites in each halo mass bin with respect to equally-massive centrals, split into star-forming, green-valley and quiescent. We discuss each stellar population parameter in the next subsections. 

\subsection{Stellar age}
The relation between light-weighted age and stellar mass (upper panels of Fig.~\ref{fig:rel_mstar_sfr}) shows a clear dependence on specific SFR: not only the mean age at fixed stellar mass shifts to older ages going from star-forming to quiescent galaxies, as expected, but also the relation is 
significantly steeper for star-forming galaxies, while it becomes progressively flatter for green-valley and for quiescent galaxies. Although the relations for satellites and central galaxies are very similar, for a given star formation activity, we do note that {\it satellite galaxies have a narrower range in stellar age at fixed mass with respect to central galaxies, showing a reduced tail toward younger ages}. We also note the tendency of the median age of satellite galaxies to be slightly older than that of central galaxies, in particular for star-forming and green-valley galaxies in the most massive halos at moderate stellar masses: although these differences are very small, they are formally significant at a level between 3 and 5$\sigma$ for SF in the mass range  $10^{10.3}-10^{10.8}M_\odot$ and at $3-7\sigma$~ for GV in the mass range $10^{10.6}-10^{10.2}M_\odot$. These are also supported by a >99\% KS probability of rejecting the null hypothesis.

The upper panels of Fig.~\ref{fig:resid_mstar_sfr} show more clearly the differences in age between satellites in different halo mass ranges and centrals. We see a small residual difference in light-weighted age of $\lesssim 0.5-1$~Gyr for star-forming and green-valley satellites residing today in the most massive halos ($M_h>10^{14}M_\odot$) with respect to equally-massive centrals. This excess in light-weighted age could indicate a reduction in the level of recent star formation activity in these galaxies compared to equally-massive central galaxies and satellites in lower-mass halos. This qualitatively reflects the result of \cite{pasquali12} where we showed that, among SDSS galaxies with a measure of gas-phase metallicity (that is, emission line galaxies), satellites have a global specific SFR lower by 0.03~dex than centrals at any given stellar mass.
This finding is also qualitatively in agreement with the detected reduction in the average specific SFR at fixed mass for star-forming satellite galaxies compared to their field counterparts found by \cite{woo17} \citep[they report a larger reduction of 0.1~dex than][]{pasquali12}. Similarly, \cite{stages_osiris17} find a 60\% suppression in the specific SFR of H$\alpha$-detected galaxies in the A901/2 cluster compared to the field.
However, we note that, contrary to what observed in Fig.~\ref{fig:rel_mstar_mhalo} and Fig.~\ref{fig:fracQ} for the population as a whole and to the studies mentioned above, once we look at galaxies with similar specific SFR the residual effect of environment on the mean light-weighted age is detected preferentially in star-forming and green-valley galaxies more massive than $10^{10}M_\odot$, which on average contain older stellar populations (upper left and middle panels of Fig.~\ref{fig:resid_mstar_sfr}). Therefore the small excess in light-weighted age of (massive) star-forming satellites may have a different origin: it may point to a star formation history more peaked at earlier times in galaxies that reside today in massive halos.

A very small ($\sim0.2-0.5$~Gyr) but systematic excess in light-weighted age is detected also for quiescent satellites in the most massive halos compared to equally-massive quiescent central galaxies (at $2-5\sigma$~significance over the mass range $10^{9.8}-10^{11.3}M_\odot$).  A similar residual dependence of age on environment was observed by \cite{Cooper10} for early-type galaxies and it was interpreted as evidence for assembly bias such that galaxies in higher-density regions formed earlier than equally-massive galaxies in less dense environments. \cite{clemens06} also found a difference of only 0.03-0.04~dex between early-type galaxies in low-density environments and those in high-density environments. 

The environment in which these massive, passive galaxies were located at the time of their quenching was likely different from the cluster-sized halos in which they are located today. Given the old ages of these quiescent galaxies ($\sim 8$~Gyr) they likely had their star formation quenched before becoming satellites of the current halo. These differences, although small (few hundred Myr), may reflect either an earlier assembly epoch for galaxies today in massive halos with respect to galaxies that reside today in lower-mass halos (assembly bias) possibly associated to a different location within the cosmic web, or the effect of environmentally-induced star formation quenching at early times in a smaller halo (pre-processing).

\subsection{Stellar metallicity}
The relation between stellar metallicity and stellar mass (middle panels of Fig.~\ref{fig:rel_mstar_sfr}) also depends on the current star formation activity, although less so than for light-weighted age. Star-forming galaxies display a steeper relation between stellar metallicity and stellar mass and they have a larger dispersion in stellar metallicity at fixed mass than galaxies with low or absent current star formation activity. The difference in stellar metallicity between quiescent and star-forming galaxies varies from 0.3~dex at $M_\ast=10^{9.5}M_\odot$~to 0.1~dex at $M_\ast=10^{10.5}M_\odot$. 
The different slopes of the mass-metallicity relation for star-forming and quiescent galaxies, and the excess metallicity of quiescent galaxies at low masses with respect to SF galaxies, have already been shown by \cite{peng15} using our stellar population catalog but a different classification into star-forming and quiescent. This result has been recently confirmed by \cite{Trussler20a} using a different set of stellar population parameters and in particular mass-weighted stellar metallicities. They interpreted the excess stellar metallicity of quiescent galaxies with respect to star-forming galaxies as an indication that `starvation' (i.e. suppression of gas inflows) is the dominant quenching mechanism at all masses, with efficient outflows being needed in addition at masses below $\rm 10^{10.2}M_\odot$. An alternative explanation has been proposed by \cite{spitoni17}: within their framework of analytical models of chemical evolution, the different slopes of the mass-metallicity relations for star-forming and quiescent galaxies require shorter gas infall timescales in quiescent galaxies leading to rapid star formation and metal enrichment at early epochs combined with stronger outflows in low-mass star-forming galaxies.
Differences between centrals and satellites and as a function of halo mass could be expected if environmental processes modulate the gas infall rate (such as strangulation and ram-pressure stripping that suppress inflows by removing the gas from the halo and from the outer disk) and/or the gas escape through outflows (e.g. galactic wind confinement).

Figure~\ref{fig:rel_mstar_sfr} shows that the dependence of the mass-metallicity relation on current star formation rate is similar for satellites and centrals: satellites have comparable stellar metallicity as central galaxies with similar specific SFR. Nevertheless, we detect a slightly higher stellar metallicity in low-mass quiescent satellites and in intermediate-mass star-forming satellites with respect to their central counterparts.
This is quantified in the middle panels of Fig.~\ref{fig:resid_mstar_sfr}, where we detect (with $3-5\sigma$~significance) an excess of $<0.1$~dex in intermediate-mass star-forming and green-valley satellites residing in $>10^{14}M_\odot$~groups. For quiescent satellites, a similar excess in stellar metallicity is detected at all stellar masses below $\sim10^{11.3}M_\odot$ and for halos more massive than $10^{13.5}M_\odot$ (with $3-8\sigma$~significance\footnote{The significance is given comparing the difference with the quadrature sum of the errors on the median. Over the quoted mass ranges a KS test rejects the null hypothesis at >98\% probability.}) and it appears to increase with decreasing stellar mass. These differences in stellar metallicity between satellites in the most massive halos and central galaxies - for either star-forming, green-valley and quiescent galaxies - display a very similar trend as stellar age with galaxy mass and halo mass, pointing to a common origin of the age and metallicity excess. Interestingly, the metallicity excess is detected only in galaxies dominated by old stellar populations, i.e. quiescent galaxies and high-mass star-forming galaxies. This suggests that the action of environment on these galaxy populations needs to be pushed back at early times.

If interpreted in terms of assembly bias, this would mean that galaxies residing today in more massive halos had a more efficient metal enrichment than galaxies in lower-mass halos. However, these differences could also be related to the amount of metals available in the star forming ISM modulated by the quenching process, i.e. a mechanism that allows star formation to continue in a metal-enriched ISM such as the suppression of inflows of metal-poor gas \citep[as discussed e.g. in ][]{pasquali12,peng15,bahe17,Maier19a}. Such mechanism would be enhanced or more effective in massive halos where satellites have, on average, an older infall time (the time when a galaxy becomes a satellite) and hence have been exposed to environmental effects for a longer time \citep{pasquali10,pasquali19}.

\subsection{Element abundance ratio \afe}
The element abundance ratio \afe~shows a strong dependence on stellar mass for quiescent and green-valley galaxies (bottom panels of Fig.~\ref{fig:rel_mstar_sfr}). Their \afe~increases by about 0.3~dex over two orders of magnitude in stellar mass. This is in agreement with the relation with stellar mass and/or velocity dispersion found by other works on early-type galaxies \citep[e.g.][]{jorgensen99,trager2000,kuntschner01,Thomas05,gallazzi06,graves09,Conroy14,labarbera14,walcher15}. 

Here we show for the first time the relation between \afe~and stellar mass for star-forming galaxies. This relation is much shallower than the one for quiescent galaxies, and the stellar populations of star-forming galaxies have median \afe~values lower than those of equally-massive quiescent galaxies. If \afe~traces star formation timescales, low \afe~values (close to solar) and a flat relation with mass are expected in galaxies in the local Universe with ongoing star formation (hence with long star formation timescales).
However the relation is not completely flat and \afe~is above solar for the 
majority of galaxies. This could be explained if the current (SFH-integrated) \afe~is set to first order by the efficiency of star formation 
in the early epoch of galaxy formation (the peak star formation rate), when enrichment of $\alpha$~elements occurs, and this efficiency is a function of galaxy final mass \citep[see][]{fontanot17}. In this case, subsequent star formation acts to dilute the early-epoch \afe~because of SN Ia products being reincorporated in stars. It must be noted that Mg and Fe lines and \mgtfe~are prominent for populations older than a few hundred Myr \citep[e.g.][]{Vazdekis15}, and therefore one may expect that our measure of \afe~mostly reflects the properties of the bulk of the (old) stellar populations. In galaxies composed by a bulge and a disk we thus expect that the inferred \afe~is mostly representative of the bulge, whose properties possibly depend on total stellar mass in a mild way, similarly to quiescent galaxies.\footnote{We also caution against over-interpreting the low-mass end of the relation for star-forming galaxies at this stage: because BC03 models follow the abundance pattern of the Milky Way, our \afe~could be underestimated at metallicities $\log(Z_\ast/Z_\odot)<-0.5$, a regime relevant for low-mass star-forming galaxies (see Appendix~\ref{A1}). However, this bias may be not very important on average given the young ages of these galaxies.}

As observed for the galaxy population as a whole in Fig.~\ref{fig:rel_mstar}, star-forming satellites and central galaxies have the same \afe, indicating that any environmentally-induced small difference in the recent star formation history of galaxies has not affected their stellar abundance pattern. Contrary to age and metallicity, this remains true even when splitting satellites according to their host halo mass (bottom-left panel of Fig.\ref{fig:rel_mstar_sfr}~and \ref{fig:resid_mstar_sfr}).
Instead, for the most massive halos ($\rm M_h>10^{14}M_\odot$ - magenta line in the middle and right bottom panels of Fig.~\ref{fig:resid_mstar_sfr}), we detect a small \afe~excess in green-valley satellites at masses $10^{10.8}-10^{11.3}M_\odot$~($4\sigma$~significance) and in quiescent satellites at masses $10^{10}-10^{10.8}M_\odot$  ($2-5\sigma$~significance\footnote{The significance is estimated comparing the difference with the error on the median. Over these mass ranges a KS test rejects the null hypothesis that the distributions are equal at >99\% confidence.}).
If interpreted in terms of star formation timescale, this points to a star formation timescale shorter in this class of galaxies with respect to equally-massive galaxies in lower density environments, but by a very small amount. By using the empirical calibration of \cite{delaRosa11} between \afe~and the time to form half of the present stellar mass we find that an excess of $0.02-0.05$~dex in \afe~would translate into a difference in half-mass time of only <$500$~Myr. Alternatively, this small excess would indicate differences in the early phases of star formation for satellites in the most massive halos, in particular larger star formation rates at early times if a SFR-dependent integrated galaxy-wide IMF is assumed \citep{fontanot17}. Interestingly, this small excess in \afe~is accompanied by a difference in light-weighted mean age of $\sim200-300$~Myr. We have checked that we find quantitatively a similar difference in mass-weighted mean age\footnote{Mass-weighted ages are estimated in the same way as light-weighted age, except for the different weighing along the SFH \citep[see][]{gallazzi08}.}, possibly suggesting that they are associated to a difference in the epoch when the bulk of stars formed rather than in the lower-redshift tail of the star formation history.

\section{Dependence on infall time}\label{sec:zones}
The extent to which environment can shape a galaxy star formation history may well depend on the time the galaxy span in that particular environment, or in other words on the infall time onto a given host environment. With this idea in mind, \cite{pasquali19} have explored the dependence of SFR and stellar population properties of satellites on the infall time onto their present-day host halo. The infall time is traced by the location ("zones") in the phase-space diagram of cluster-centric velocity and cluster-centric distance. In particular \cite{pasquali19} use hydrodynamical zoom-in simulations of galaxy clusters to calibrate the dependence on infall time and define zones in phase-space with different mean infall time that minimize the scatter in infall time within each zone. The time of infall is here defined as the time a satellite first crosses the virial radius of the main progenitor of its present-day host halo.

\begin{figure}
\centerline{\includegraphics[width=9truecm]{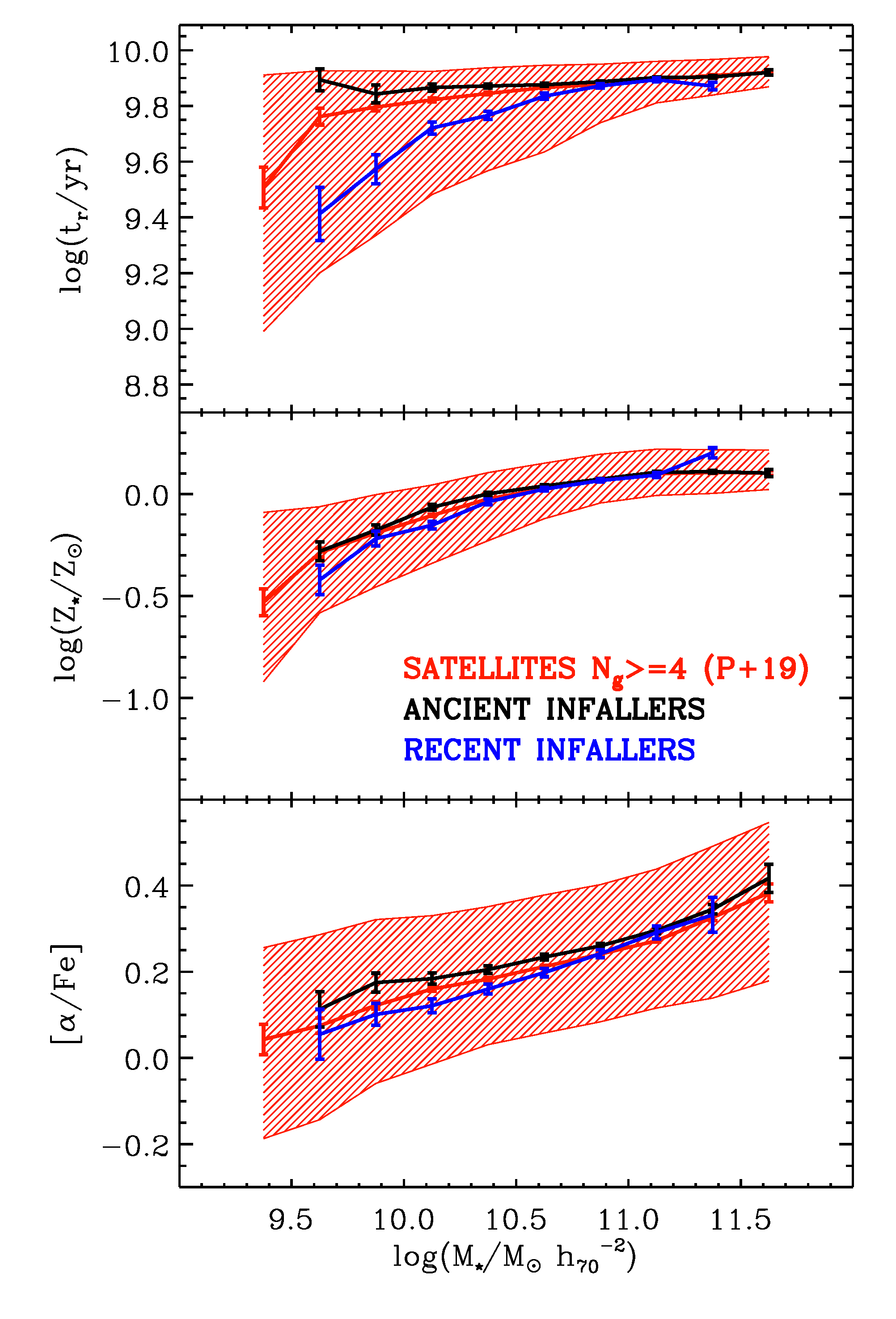}}
\caption{Relations between r-band mean stellar age (top panel), stellar metallicity (middle panel), \afe~(bottom panel) and stellar mass for satellite galaxies in groups with at least four members \citep[i.e. the sample used in][]{pasquali19}. The solid red line with error bars traces the median trend, while the hatched region encloses the 84th-16th interpercentile range. The black and blue lines with error bars trace the median trend for satellites classified as ancient infallers and as recent infallers, respectively, on the basis of their location in phase space \citep[see][]{smith19}. The error bars indicate the error on the median.}
\label{fig:rel_mstar_zones}
\end{figure}

We explore whether the differences between centrals and satellites presented in the previous sections depend on the infall time. We restrict the analysis to the satellites used in \cite{pasquali19}, i.e. those residing in groups with at least four members. Following \cite{smith19}, we distinguish between `ancient infallers' (with an average infall time older than 5 Gyr, corresponding to zone number smaller than 2; see their Fig. 4) and `recent infallers' (with an average infall time younger than 2.5 Gyr, corresponding to zone number larger than 5). Fig.~\ref{fig:rel_mstar_zones} shows the global stellar population scaling relations as in Fig.~\ref{fig:rel_mstar} but only for these subsets of ancient and recent infaller satellites. We note that the mean relations for this subsample of satellites are shifted to older stellar ages and higher metallicities than the whole satellite sample shown in Fig.~\ref{fig:rel_mstar}: this is a consequence of the cut in number of group members ($> 4$), which leads to excluding the least massive halos, and of the dependence of stellar population properties on host halo mass. 

The important result of Fig.~\ref{fig:rel_mstar_zones} is that ancient infallers follow different relations than recent infallers. In particular they are significantly older and slightly metal-richer\footnote{The difference in median metallicity is significant, compared to the error on the median, only for few mass bins. However the distributions in metallicity for ancient and recent infallers are different with a KS probability of $>98\%$~at masses $<10^{10.6}M_\odot$.} than recent infallers of similar stellar mass. It is worth noting that ancient infallers have on average light-weighted mean ages of 8-9 Gyr, hence older than the mean epoch at which they became satellites on their current halo. The extent to which the ages and metallicities of ancient and recent infallers differ is similar to that by which the ages and metallicities of satellites residing in different halo masses differ (see Fig.~\ref{fig:rel_mstar_mhalo}). Interestingly, we also see that over the mass range $10^{9.9}-10^{10.6}M_\odot$ ancient infallers have $0.04-0.07$~dex larger \afe~at fixed mass with respect to recent infallers at $2-3\sigma$~significance. This is a different representation of the results presented in \cite{pasquali19} on the trends of sSFR, age, metallicity and \afe~with infall time for satellites in different bins of stellar and halo mass, which are stronger for satellites with mass lower than $10^{10.5}M_\odot$ and in halos more massive than $10^{13.5}M_\odot$. This indicates that time of infall is an important factor in shaping satellites' star formation history and it is suggestive of a signature of an early (even prior to infall onto the current halo) and more prolonged exposure to environmental processes by ancient infallers.

\begin{figure*}
\centerline{\includegraphics[width=15truecm]{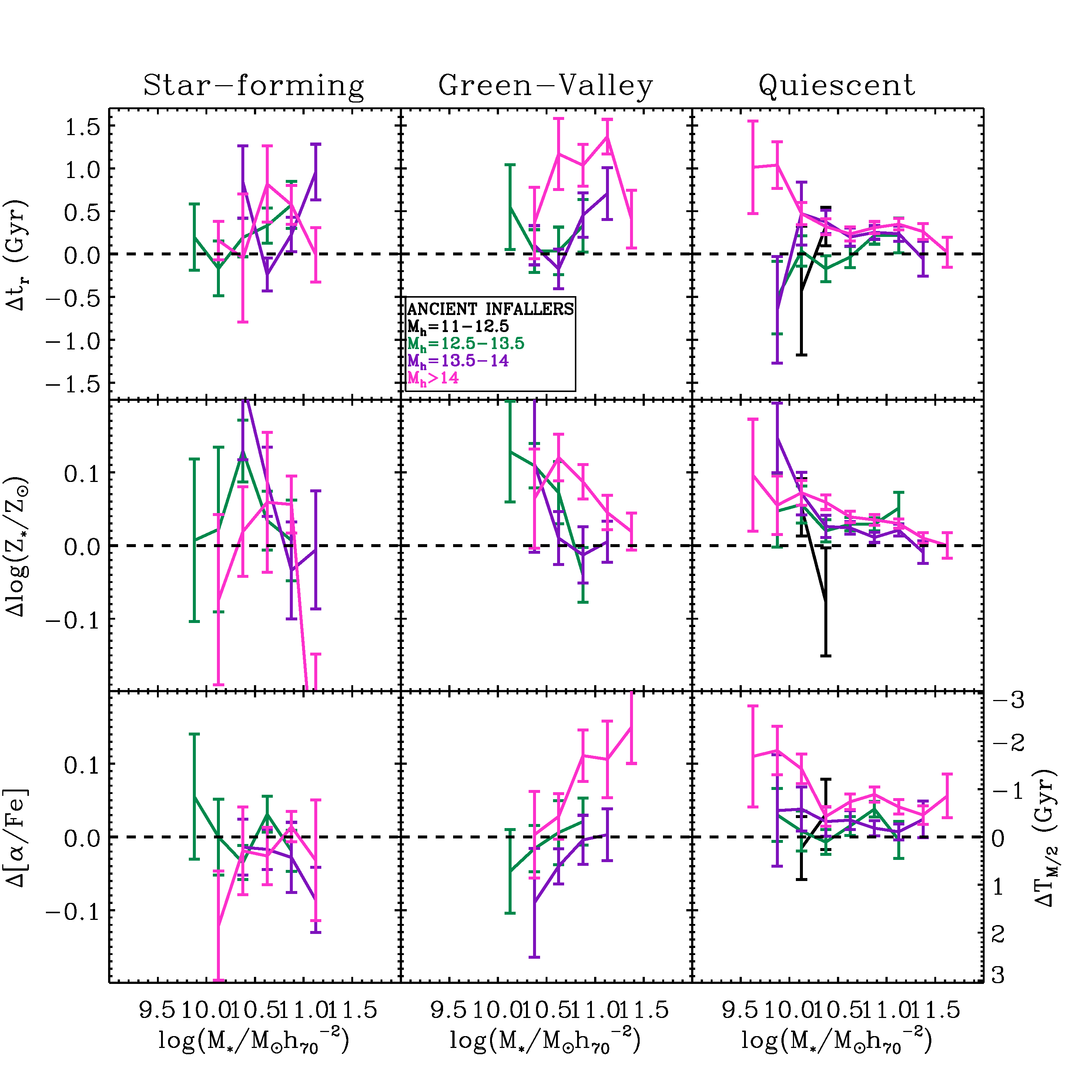}}
\caption{Difference in light-weighted age (upper panels), stellar metallicity (middle panels), \afe~(bottom panels) between satellites classified as ancient infallers, divided into bins of halo mass as in Fig.\ref{fig:resid_mstar_sfr} (colored lines), and central galaxies at fixed stellar mass. The error bars indicate the quadrature sum of the errors on the median for satellites and centrals. Galaxies are divided into star forming (left column), green-valley (central column) and quiescent (right column). Only bins with at least 10 galaxies are shown.}
\label{fig:Dparam_mstar_mhalo_sfr_ancient}
\end{figure*}

\begin{figure*}
\centerline{\includegraphics[width=15truecm]{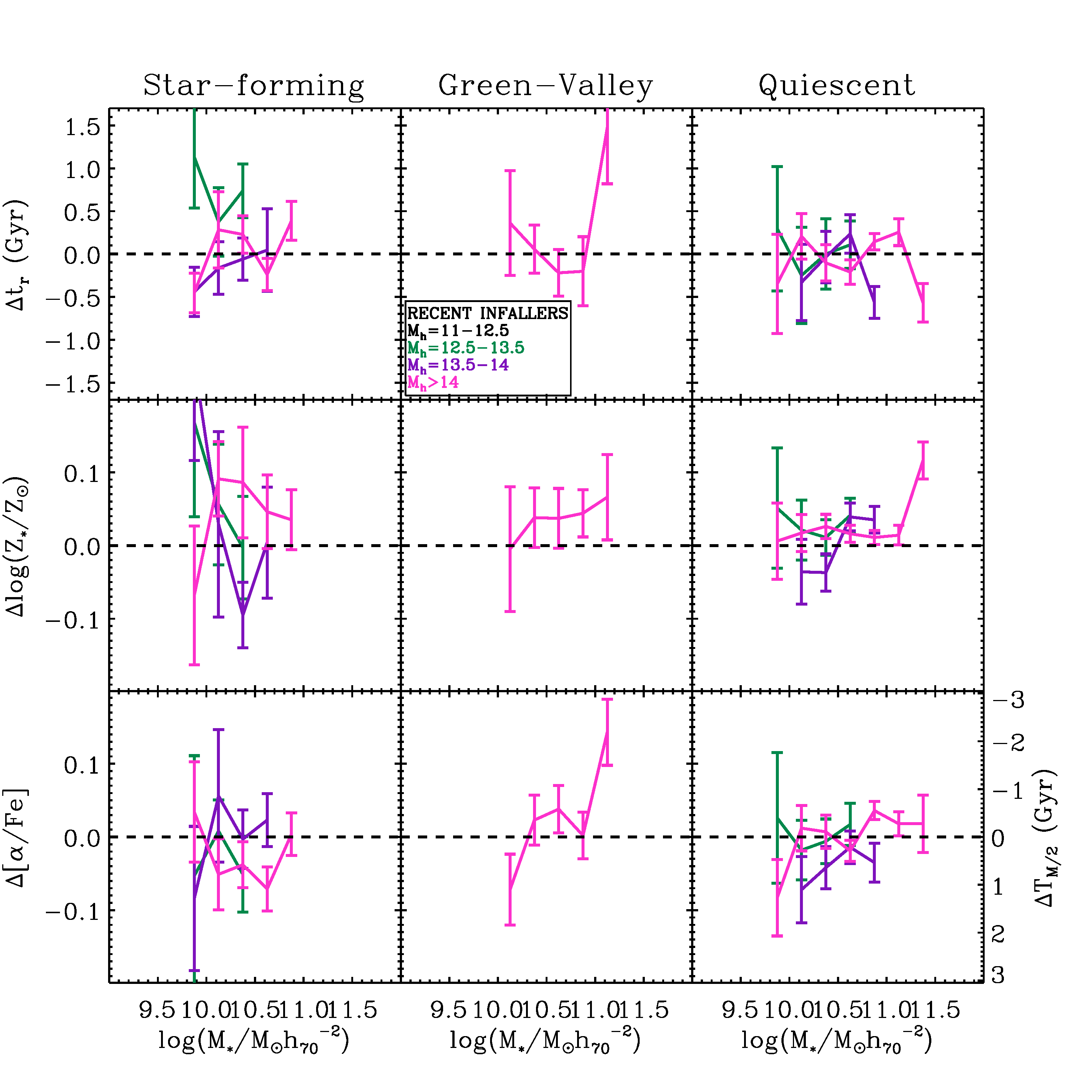}}
\caption{Difference in light-weighted age (upper panels), stellar metallicity (middle panels), \afe~(bottom panels) between satellites classified as recent infallers, divided into bins of halo mass as in Fig.\ref{fig:resid_mstar_sfr} (colored lines), and central galaxies at fixed stellar mass. The error bars indicate the quadrature sum of the errors on the median for satellites and centrals. Galaxies are divided into star forming (left column), green-valley (central column) and quiescent (right column). Only bins with at least 10 galaxies are shown.}
\label{fig:Dparam_mstar_mhalo_sfr_recent}
\end{figure*}

We explore in more detail any signature of a prolonged exposure to environmental effects on the stellar populations in Figg.~\ref{fig:Dparam_mstar_mhalo_sfr_ancient}~and~\ref{fig:Dparam_mstar_mhalo_sfr_recent} separately for ancient and recent infallers, respectively. Similar to Fig.~\ref{fig:resid_mstar_sfr}, we remove the main effect due to the varying fraction of quiescent galaxies by dividing satellites into star-forming, green-valley and quiescent. For each subsample we plot the difference in age (upper panels), stellar metallicity (middle panels) and \afe~(lower panels) between satellites residing in halos of different mass and equally-massive central galaxies. Comparing Fig.~\ref{fig:Dparam_mstar_mhalo_sfr_ancient}~(ancient infallers) and Fig.~\ref{fig:Dparam_mstar_mhalo_sfr_recent}~(recent infallers) with Fig.~\ref{fig:resid_mstar_sfr}~it is clear that {\it the excess in age, metallicity and \afe~detected for satellites in massive halos with respect to centrals is largely driven by the ancient infallers satellite population}. For recent infallers (Fig.~\ref{fig:Dparam_mstar_mhalo_sfr_recent}) we do not detect any significant and systematic difference in their stellar populations with respect to centrals, regardless of their current star formation activity and of the mass of their current halo. On the contrary, ancient infallers (Fig.~\ref{fig:Dparam_mstar_mhalo_sfr_ancient}) with low star formation activity (quiescent and green-valley) display an excess in mean stellar age, stellar metallicity and \afe~with respect to their central counterparts. For passive galaxies this is evident already for halo masses larger than $10^{13.5}$M$_\odot$.
In particular, we find: i) an excess in light-weighted age of $<1$~Gyr for massive ($10^{10.6}-10^{11.1}M_\odot$) green-valley ancient infallers in halos larger than $\rm 10^{14}M_\odot$, and of $0.3-1$~Gyr for passive ancient infallers in the mass range $10^{10.3}-10^{11.1}M_\odot$ for halo masses  $\rm >10^{13.5}M_\odot$ and at all stellar masses for halos larger than $\rm 10^{14}M_\odot$ , ii) an excess in stellar metallicity of $<0.1$~dex for green-valley and quiescent ancient infallers in halos more massive than $10^{14}M_\odot$ at masses $10^{10.1}-10^{11.1}M_\odot$ , iii) an excess in \afe~of 0.1~dex for green-valley (at masses $>10^{10.8}M_\odot$) and quiescent ancient infallers (at virtually all masses for halos more massive than $10^{14}M_\odot$).\footnote{These are the mass ranges over which ancient infallers differ from centrals at more than $3\sigma$~significance and with a KS probability >99\%.}
These results are consistent with \cite{pasquali19} and with \cite{smith19} who find that ancient infallers reached 90\% of their stellar mass ~2 Gyr earlier than recent infallers. We also confirm the power of distinguishing galaxies according to their phase-space location (or infall time) and according to their current star formation activity to detect signature of environmental effects in halos with masses down to $10^{13.5}M_\odot$. 

Galaxies that only recently ($\sim1.5$~Gyr ago on average) fell onto (the main progenitor of) their present-day halo do not appear to have (yet) their stellar populations altered by their current environment (Fig.~\ref{fig:Dparam_mstar_mhalo_sfr_recent}). The fact that they have similar stellar populations as equally-massive central galaxies also suggests that they were not affected by the environment prior to becoming satellites of their current halo: this means that if they were already satellites in a smaller halo this didn't affect their star formation history, i.e. that pre-processing has not been important, or that they were located in a low-density environment similar to the field. In \cite{pasquali19} satellites located at high zone numbers (recent infall epochs) are found to be slightly older and metal-richer than the average field galaxy, at fixed stellar and halo mass. This result was obtained without distinguishing star-forming and passive galaxies, and was interpreted as a possible sign of pre-processing. In light of our analysis, it can be justified as due to a higher fraction of quiescent galaxies among recent infallers with respect to field galaxies. It is then these recently accreted quiescent satellites to have experienced pre-processing. We checked that the fraction of quiescent galaxies among recent infallers varies from 27\% for halos less massive than $10^{12.5}M_\odot$ to 70\% for halos more massive than $10^{14}M_\odot$, compared to 40\% for centrals in halos less massive than $10^{12.5}M_\odot$ (possibly isolated galaxies).

\section{Summary and Discussion}\label{sec:conclusions}
We have investigated the imprint of environmental effects on galaxy evolution left on the present-day stellar populations in galaxies according to their present-day environment. Specifically, we have compared the properties (light-weighted age, stellar metallicity, \afe) of the stellar populations in a sample of 26023 satellite galaxies to those in a sample of 87284 central galaxies, making use of the SDSS-DR7 \cite{wang14} group catalog. In addition to stellar metallicity and age derived as in \cite{gallazzi05}, in this work we also derive estimates of \afe~for galaxies with any star formation activity.

\subsection{`Nurture' and delayed-then-rapid quenching}
We contrast the properties of satellite galaxies to those of equally-massive central galaxies, controlling also for host halo mass.
We find that:
\begin{itemize}
\item For high mass galaxies ($M_\ast\gtrsim10^{10.8}M_\odot$) satellites and centrals have on average the same light-weighted age and stellar metallicity. These properties increase only mildly with stellar mass and have a negligible dependence on halo mass for satellite galaxies (Figg.~\ref{fig:rel_mstar} and~\ref{fig:rel_mstar_mhalo}).
\item At lower stellar masses satellite galaxies have median light-weighted ages older by $\sim0.3-0.4$~dex and median stellar metallicities higher by $\sim0.1$~dex than equally-massive central galaxies. We also note that not only the median age and metallicity differ but also the overall distributions at fixed mass, indicating a dearth of young, metal-poor galaxies and an excess of old, metal-rich galaxies among the satellites population (Fig.~\ref{fig:rel_mstar}). The differences in median light-weighted age and, to a lesser degree, in median stellar metallicity between satellites and equally-massive central galaxies increase with increasing host halo mass (Fig.~\ref{fig:rel_mstar_mhalo}). These results are in agreement with what found in \cite{pasquali10} for SDSS DR2. 
\item The \afe~abundance ratio increases monotonically by almost 0.4~dex over three orders of magnitude in stellar mass. Both the median trend and the $\rm16^{th}-84^{th}$ percentile range are remarkably similar for satellite and central galaxies, indicating that the \afe~is primarily set by the galaxy stellar mass (Fig.~\ref{fig:rel_mstar}). 
\item The comparison of centrals and satellites by controlling for both stellar mass and halo mass is hampered by the small overlap range in masses. In a given bin of halo mass, satellites and centrals appear to follow similar scaling relations with stellar mass but covering different mass ranges. This could  argue against satellite-specific processes shaping satellites' stellar populations \citep{wang18}. However, we notice two results which we think are evidence of environment acting specifically on satellite galaxies: i) for $\log M_h/M_\odot<12.5$ satellites are on average slightly older and more metal enhanced than centrals at fixed $M_\ast$, as a result of their slightly higher fraction of quiescent galaxies; ii) old and metal-rich galaxies are only found either at high stellar masses or among the satellite population in massive halos at intermediate to low stellar masses.
\item We have shown that the different scaling relations between centrals and satellites are to first order the result of the different mass-dependent fraction of quiescent galaxies among centrals and satellites (Fig.\ref{fig:fracQ}, see also Wetzel et al 2013). We explore the age, metallicity and, for the first time, \afe~scaling relations with stellar mass for satellites and centrals with similar specific SFR, finding them to be very similar (Fig.\ref{fig:rel_mstar_sfr}). This suggests at face value similar past star formation and metal enrichment histories for satellites and central galaxies at fixed mass and current SF activity. The same result has been recently found with the independent analysis of \cite{Trussler20b}.
\end{itemize}

The results above indicate that the stellar populations in present-day galaxies retain significant information about the environment-driven quenching of star formation for galaxies less massive than $\sim3\times10^{10}M_\odot$. The differences in age and their increase with host halo mass reveals that halos of present-day high mass have facilitated the process of quenching, especially in low-mass galaxies. When and how fast does environment affect the star formation history of galaxies as witnessed by their present-day stellar populations? The main signature of environmental effects is the higher fraction of quiescent galaxies among satellites and with increasing halo mass, but typically star-forming galaxies in clusters are found to have similar levels of star formation to those in the field \citep[e.g.][]{balogh04,poggianti08}. This, together with the fact that we find the stellar population scaling relations of centrals and satellites to be virtually identical when controlling for the varying quiescent fraction \citep[see also][] {Trussler20b}, would argue for quenching happening on a rather short timescale: galaxies are either `on' or `off' and there is not a gradual transition in the stellar population properties. If this were the case, we might expect to see a signature of fast environmental quenching in the \afe~being higher for satellite galaxies, if we assume that \afe~is a tracer of star formation timescale and if quenching happens early on as soon as a galaxy becomes a satellite. On the contrary, we find that to first order \afe~scales with stellar mass in the same way for central and satellite galaxies. This would suggest that the timescale of star formation is set to first order by the galaxy mass (internal processes) and that it is not affected by environment. Or, in other words, it indicates that any environmentally driven star formation quenching happens significantly after the bulk of star formation and metal enrichment occurs, on an overall timescale of few or several Gyr to allow SNIa products to be incorporated in stars. 

These two apparently contradicting results can be reconciled in a two-phase quenching scenario, also referred to as `delayed-then-rapid quenching': star formation would continue for a few Gyr when a galaxy is accreted onto a halo before being rapidly quenched because of the accumulated effect of environmental processes altering the gas content and distribution within the galaxy. This scenario was first proposed by \cite{wetzel13} who constrained the delay time to vary between 5 and 2 Gyr decreasing with increasing stellar mass and the time of actual quenching to be 0.2-0.8 Gyr. Similarly long overall timescales (not necessarily in two phases), between 3 and 6 Gyr, for galaxies to quench after becoming satellites have been also determined based on comparison of the observed passive fractions with predictions from SAM \citep[e.g.][]{delucia12,hirschmann14} and from N-body cosmological simulations \citep[e.g][]{oman16}. Finally, in \cite{pasquali19} we determine that it takes between 4 and 6 Gyr before the specific SFR of $\log M_\ast<10.5$ satellites decreases below $10^{-11}M_\odot/yr$ and $\sim2-3$~Gyr for higher mass satellites. This might explain the similar \afe~of satellites and centrals that we find in this work.

In the context of delayed-then-rapid quenching, the higher stellar metallicities of satellite galaxies can be understood if star formation continues in a more metal-rich ISM with respect to centrals. Several mechanisms acting on satellites can cause the ISM to reach higher metallicities. Removal of gas from the halo (`strangulation', acting on a few Gyr timescale) and then from the outer disk (ram-pressure stripping, acting on 1-2 Gyr timescale) \citep{BoselliGavazzi06} would inhibit inflows of metal-poor gas toward the central regions of galaxies, thus allowing the subsequent generations of stars to reach higher metallicities. The combined action of strangulation and ram-pressure stripping could produce the two-phase quenching: on long timescales when only the hot gas reservoir is removed and then on rapid timescales when the cold gas is stripped via ram-pressure. It has been suggested already by previous works in order to explain the differences observed in gas-phase metallicity between satellites and centrals \citep{pasquali12, bahe17} as well as the enhanced gas-phase metallicities of galaxies located in the central regions of both low and intermediate redshift clusters \citep{Maier19a,Maier19b}. Alternatively, higher gas-phase metallicities (and, subsequently, higher stellar metallicities) could result from the confinement of galactic winds under the pressure of the hot ICM \citep{Mulchaey10}, preventing outflows to remove metals from the ISM \citep{pasquali12}. However, based on cosmological hydrodynamic simulations and analytic models \citep{bahe12}, this process is expected to be sub-dominant with respect to stripping of the hot gas or to operate only when galaxies have already lost their hot gas.
Gradual stripping of the hot gas reservoir appears as a key ingredient to be treated in semi-analytic models of galaxy formation in order to reproduce the observed quenched fraction in central and satellite galaxies and the long timescales of satellite quenching inferred from observations \citep{xie20,cora18}. However discrepancies between both semi-analytic models and hydro-dynamical simulations and observations remain for low-mass galaxies in massive halos and high-mass galaxies in low-mass halos which are not trivial to approach \citep[see discussion in][]{xie20}.

\subsection{`Nature' and early environmental effects}
We have explored in detail differences in the present-day stellar populations of satellites and centrals, accounting for the varying quiescent fraction in the two samples. Controlling for halo mass and for infall time, based on the phase-space location of galaxies, small but interesting differences emerge for relatively massive galaxies. 
\begin{itemize}
\item We find a small excess in light-weighted age and stellar metallicity for intermediate-mass star-forming satellites residing in the most massive halos ($\gtrsim10^{14}M_\odot$; upper and middle panels of Fig.~\ref{fig:resid_mstar_sfr}). At face value, this may suggest a reduction in the star formation rate of satellites in cluster-sized halos, possibly due to the suppression of metal-poor gas supply that would act to increase the observed stellar metallicity along with aging of the stellar populations. If this were the case the small environmental-induced SFR reduction should not affect the \afe~of star-forming satellites with respect to centrals (lower panel of Fig.~\ref{fig:resid_mstar_sfr}): the continued star formation activity, and the consequent recycling of Type Ia SN products, would decrease \afe~in a similar way as for equally-massive centrals. Few studies have identified small levels of reduction in SFR in low-mass ($<10^{10}M_\odot$) cluster galaxies \citep[e.g.][]{woo17,pasquali12}. However, the differences in age and metallicity that we observe for star-forming galaxies are not detected at a significant level in low-mass galaxies, but manifest mainly at intermediate masses (between $10^{10}$ and $10^{11}M_\odot$), i.e. in galaxies with light-weighted ages between 3 and 5 Gyr. Interestingly, these intermediate-mass star-forming galaxies may resemble the population of dusty star-forming galaxies with reduced SFR and with preferentially intermediate mass values found in low-redshift clusters \citep[e.g][]{wolf09,gallazzi09,stages_osiris17}. Alternatively, what we are witnessing here is not a reduction in SFR but rather earlier formation epochs (associated to the earlier collapse of more massive halos) and higher metal enrichment efficiencies but similar formation timescales (if \afe~traces SF timescale) in star-forming satellites of massive halos.

\item For passive galaxies we find that satellites in halos more massive than $10^{14}M_\odot$ show very small but systematic excess in light-weighted age (few hundred Myr), stellar metallicity ($<0.1$~dex) and \afe~($\lesssim0.05$~dex) over almost the whole mass range probed by our sample ($10^{9.8}-10^{11.3}M_\odot$) compared to satellites in lower-mass halos and central galaxies. Rejuvenation of low-mass ($<10^{10.8}M_\odot$) quiescent satellites with younger ages and lower \afe~(but higher stellar metallicity) in low-density environment, as observed by \cite{Thomas10}, does not provide an explanation for the differences we observe. 
Instead, our results are in line with the relation found by \cite{Cooper10} between light-weighted age and environmental density, whereby galaxies with older stellar populations prefer higher-density regions with respect to galaxies with similar color and luminosity. This holds also after removing galaxies with residual recent SF, i.e. the effect is not driven by a larger spread toward younger ages in lower-density environment as instead found by other works \citep{gallazzi06,Thomas10}, but a shift in the old population. This was interpreted in \cite{Cooper10} as evidence of assembly bias with galaxies located in higher density regions forming earlier. Similarly, a difference in age of $\sim0.05$~dex per decade in local density was inferred by a Fundamental Plane analysis of early-type galaxies in \cite{labarbera10}, in agreement with what we observe in Fig.~\ref{fig:resid_mstar_sfr} \citep[see also][]{bernardi06}. In line with our results, \cite{mcdermid15} found that, among $\rm ATLAS^{3D}$ galaxies, Virgo early-type galaxies are 2.5~Gyr older and more $\alpha$-enhanced than field early-type galaxies: based on the fact that such trends are observed not only on light-weighted but also on mass-weighted properties and on the old ages of these galaxies, they argue that these trends cannot be reproduced by only suppressing the recent star formation, but they are a manifestation of a higher efficiency of early star formation in today's cluster galaxies.

\item The stellar populations of satellite galaxies not only depend on the group halo mass but also on the infall time. We distinguish `ancient infaller' and `recent infaller' satellites according to their location in the phase-space diagram, calibrated onto infall time \citep{pasquali19}. {\it We find marked differences in the scaling relations of ancient and recent infallers, the first being older, more metal-rich and more $\alpha$-enhanced than the latter}. By separating recent and ancient infallers and confronting satellites and centrals with similar current star formation activity, we do not detect any significant systematic difference for recent infallers, suggesting that their stellar populations have not been affected yet by their current environment (Fig.~\ref{fig:Dparam_mstar_mhalo_sfr_recent}). Instead we see that {\it the differences discussed for the general satellite populations pertain only to ancient infallers in high-mass halos ($\geq10^{14}M_\odot$)}, i.e. satellites that accreted more than 5 Gyr ago onto their current halo (Fig.~\ref{fig:Dparam_mstar_mhalo_sfr_ancient}). We also note that galaxies that reside today in massive halos (more massive than $\sim10^{13.5}M_\odot$) are expected to have, on average, high redshift of infall \citep[$z\gtrsim2$, see][]{pasquali10} onto halos less massive than the present-day halo. 
\end{itemize}

Given the old ages and the old infall times of the satellites affected by the observed excess in age, metallicity and \afe, it is likely that quenching occurred $\gtrsim8$~Gyr ago ($z\gtrsim1$) in a smaller-mass halo than the current one. It is thus not directly associated to cluster-specific mechanisms acting on galaxies falling into large halos. We argue instead that the slightly older ages, higher stellar metallicities and \afe~observed at $z\sim0$~for ancient infaller satellites in massive halos reflect differences in their early star formation history, with more efficient star formation and metal enrichment in the progenitors of present-day galaxies that were located close to the cosmic density peaks, possibly aided by pre-processing in group-sized halos.
If \afe~traces SF timescales, the observed differences point to timescales shorter by a few hundred Myr in quiescent satellites compared to centrals. 
On the other hand the \afe~excess may be a consequence of the higher star-formation activity for the progenitors of satellite galaxies if a SFR-dependent IMF is allowed.

\subsection{Concluding remarks}
We have highlighted the importance of controlling for galaxy star-formation activity, halo mass and infall time to disentangle the subtle effects of `environment' on the past star formation and metal enrichment histories of galaxies, as witnessed by their present-day stellar populations. {\it Environment manifests both through `nurture' effects in relatively low-mass galaxies, altering the star formation history on globally long timescales, and `nature' effects acting on the early phases of the formation of galaxies located close to density peaks}. It is possible to observe such effects only thanks to high-S/N continuum spectroscopy and the large statistic of surveys such as SDSS. 

One should keep in mind that scaling relations are not evolutionary sequences of individual galaxies and that comparing centrals and satellites at fixed final mass does not mean comparing satellites with their likely progenitors. The stellar mass of a galaxy is intrinsically connected with the star formation and assembly history and hence evolves with time. What we can do observationally is comparing the end-point of the evolution of centrals and satellites: a satellite galaxy and a central galaxy that reached today the same stellar mass have done so following a different evolutionary track that put them on average onto different stellar population scaling relations. Ideally one would like to compare galaxies with the same mass at redshift of infall and track their differential evolution. In fact, part of the differences between centrals and satellites at fixed {\it present-day} stellar mass may be due to satellites having stopped their growth because of environmental effects. In other words, for a given progenitor mass, a central would end up with a larger stellar mass than a satellite. If this is the case, satellites would be better compared to centrals of larger present-day stellar mass, but this connection is hard to establish from the data alone.

It is paramount to push such studies at higher redshift, where the younger Universe age allows for a finer resolution in the early star formation histories of galaxies, following galaxy populations in their changing environment, and tracing the evolution with redshift of stellar population scaling relations to approach the analysis from the `starting point' and better connect progenitors and descendants.

\section*{Acknowledgements}
We thank the referee for the constructive report.
We are grateful to Gabriella De Lucia and Fabio Fontanot for a careful reading of the manuscript and for insightful comments.
AG and SZ acknowledge support by the Istituto Nazionale di Astrofisica (PRIN-SKA 2017 program 1.05.01.88.04). S.Z. has been supported by the EU Marie Curie Career Integration Grant "SteMaGE" Nr. PCIG12-GA-2012-326466  (Call Identifier: FP7-PEOPLE-2012 CIG).

\section*{Data availability}
The SDSS DR7 group catalog is available at http://gax.shao.ac.cn/data/Group.html.
The SFR and emission line properties of SDSS DR7 main galaxy sample used in this work are available at https://wwwmpa.mpa-garching.mpg.de/SDSS/DR7. 
The stellar population catalog for SDSS DR7 main galaxy sample is available upon reasonable request to the corresponding author (ages and stellar metallicities of previous releases are available at https://wwwmpa.mpa-garching.mpg.de/SDSS/).

\bibliographystyle{mnras}
%\bibliography{paper}

\appendix
\section{\afe~abundance ratio: observational diagnostic and calibration}\label{A1}
\begin{figure}
\centerline{\includegraphics[width=9truecm]{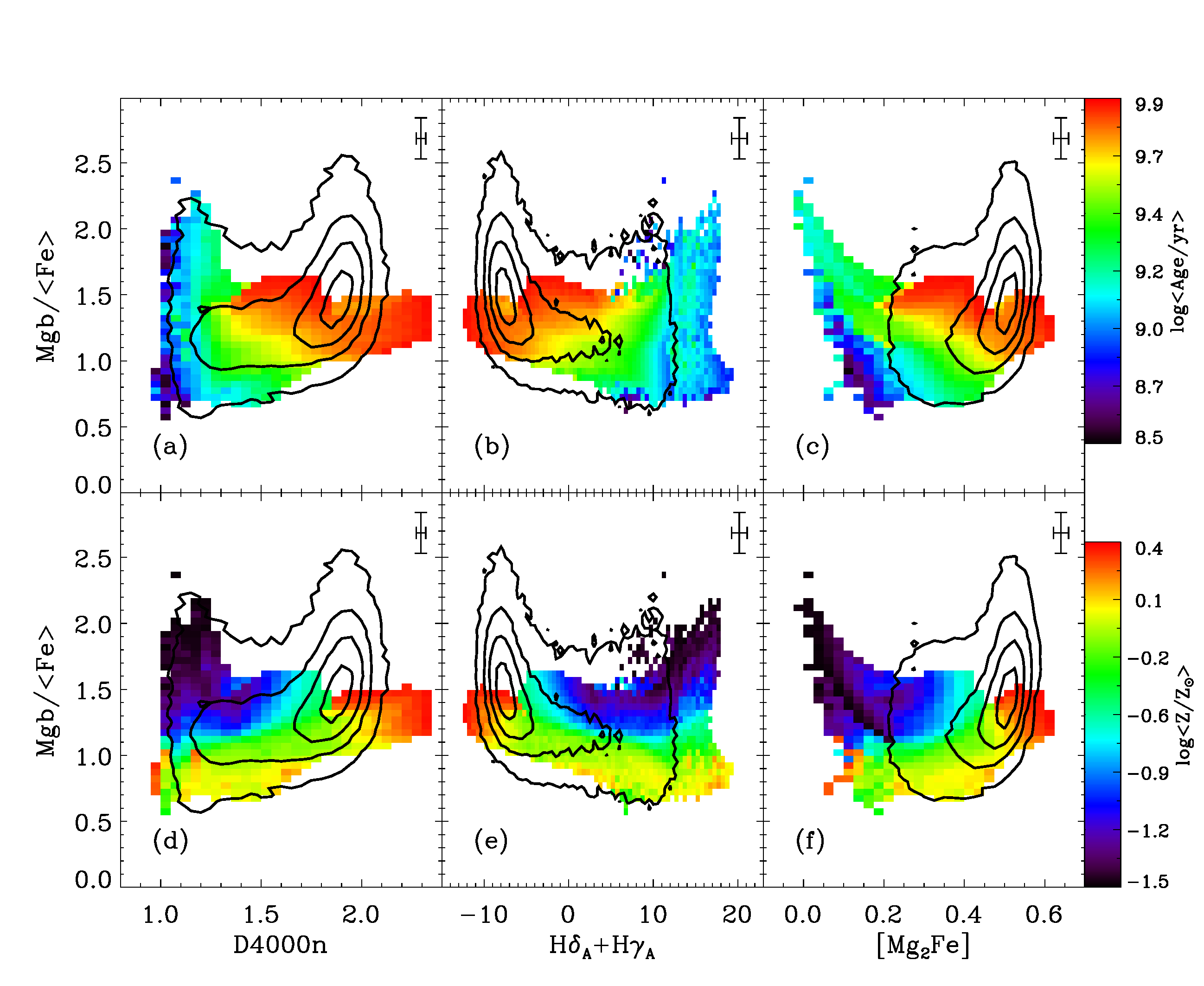}}
\caption{The \afe-sensitive absorption index \mgtfe~as a function of the age- and metallicity-sensitve indices \dn, \hdg, \mgtwofe. The distribution of the galaxies in our sample (black contours) is overplotted on top of the distribution of the models in our library, color-coded according to the mean light-weighted age (upper panels) and the mean metallicity (lower panels) in each index-index bin.}\label{fig:Dmgfe_indx_plane}
\end{figure}

\begin{figure}
\centerline{\includegraphics[width=9truecm]{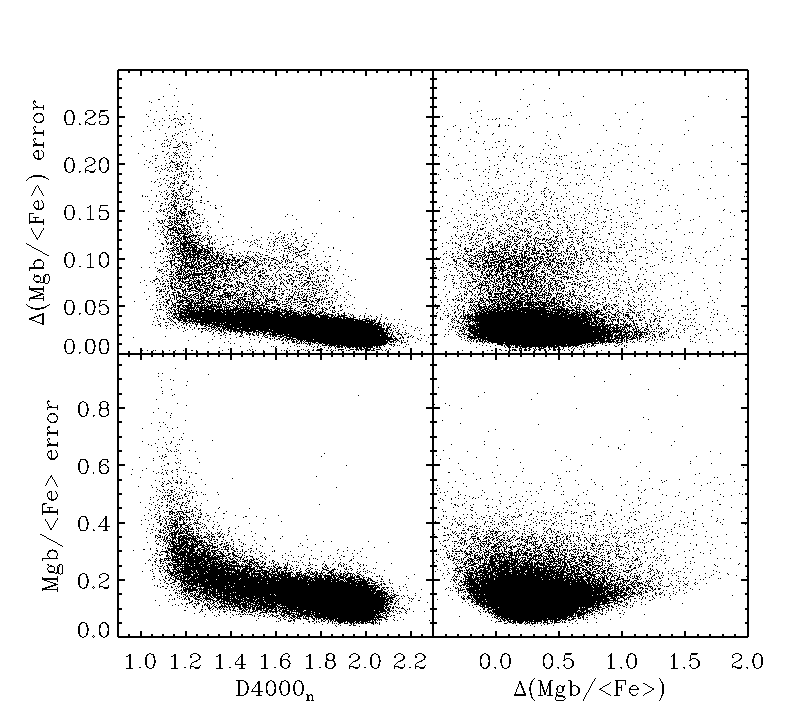}}
\caption{Observational error on the index ratio \mgtfe~as a function of \dn~(lower left panel) and of the \afe~empirical estimator \Dmgfe~(lower right panel). This can be compared with the uncertainty due to model degeneracies estimated as half of the $\rm 84^{th}-16^{th}$ percentile range of the PDF of \Dmgfe (upper panels).}\label{fig:Dmgfe_error_indx}
\end{figure}

The distribution in the \afe-sensitive index \mgtfe~is shown in Fig.~\ref{fig:Dmgfe_indx_plane} as a function of the age- and metallicity-sensitive features \dn, \hdg, \mgtwofe~used to estimate light-weighted age and metallicity. The contours for the full sample of 113307 high-S/N SDSS DR7 galaxies are overlaid on top of the distribution of the models in our library (Sec.~\ref{sec:age_met}), which is color-coded according to the average light-weighted age (upper panels) or the average stellar metallicity (lower panels) in index-index bins. While at low \dn~and high \hdg~the models cover the index range of the data, at high \dn~values a significant fraction of the observed galaxies have \mgtfe~higher than our solar-scaled models, indicative of non-solar \afe~abundance ratio. The error on \Dmgfe~include the contribution from the observational uncertainty on \mgtfe~and from the range of allowed models as expressed by the $16th-84th$ interpercentile half-range of the PDF of \Dmgfe. The two components are compared in Fig.\ref{fig:Dmgfe_error_indx}. For galaxies with high \dn~values the PDF of \Dmgfe~(upper panels) is very narrow, while for younger galaxies (lower \dn) the models have a larger range of \mgtfe~resulting in larger PDF of \Dmgfe. The narrow PDFs of \Dmgfe~are also the result of the negligible dependence of \Dmgfe~on age and metallicity, which motivates the choice of \Dmgfe~as \afe~indicator. The observational uncertainty on \mgtfe~is much larger and dominates the error budget on \Dmgfe.

\begin{figure*}
\centerline{\includegraphics[width=15truecm]{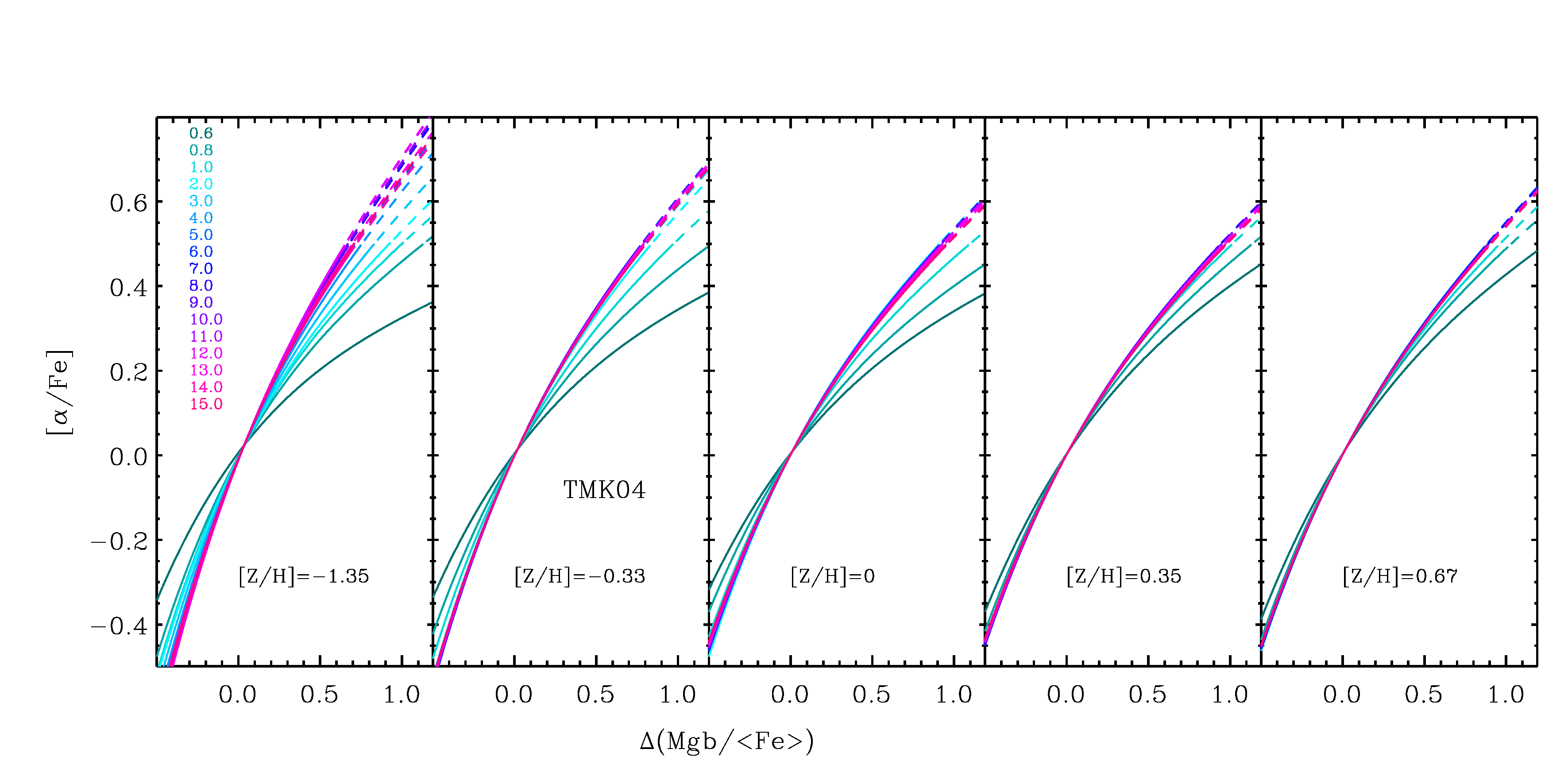}}
\caption{Calibration between \Dmgfe~and \afe~based on the TMK04 SSP models for a range of ages (colored lines; values reported are SSP ages in Gyr) and metallicities (indicated in each panel). Dashed lines indicate the range where the relations have been linearly extrapolated.}\label{fig:afe_calib_TMK04}
\end{figure*}

We calibrate the relation between \Dmgfe~and \afe~using population synthesis models of SSPs with variable \afe. We adopt as default the TMK04 models (see Sec.~\ref{sec:afe}). In Fig.~\ref{fig:afe_calib_TMK04} we show the relations between \afe~and \Dmgfe~for TMK04 models of different ages (colored lines) and of different metallicity (indicated in each panel). We fit a third order polynomial and extrapolate the relation in the range not covered by the models (dashed lines). Remarkably, these relations are virtually indistinguishable for ages older than 1~Gyr and metallicities of [Z/H]=$-0.33$ or more. For the highest metallicity the variation with age are even smaller. 

In order to check the sensitivity of the \afe~estimates on the models used to calibrate the \Dmgfe-\afe~relation we have repeated the procedure described in Sec.~\ref{sec:afe} with other four sets of models: i) the \cite{TMJ11} models spanning the same age, metallicity and \afe~range as our default choice; ii) the differential models based on the \cite{coelho07} theoretical models calibrated either on the \cite{bc03} models or the MILES v9.1 models \citep{miles11} \citep[see][]{walcher09,walcher15}, spanning the age range between 3 and 12~Gyr and between 2 and 13~Gyr respectively, $[\rm Fe/H]=-0.5,-0.25,0.,0.2$ and \afe$=0,0.2,0.4$; iii) the $\alpha$-MILES models of \cite{Vazdekis15}, with age varying between 0.5 and 14 Gyr, metallicity between [Z/H]=-1.26 and 0.26 and $\afe=0, 0.4$. We find very similar relations between \Dmgfe~and \afe~and a negligible dependence on age and metallicity or Fe abundance for all these models for ages older than 1~Gyr and metallicities [Z/H]$>1$ (\afe~can vary by at most $0.05-0.1$~dex for younger ages and lower metallicities). We prefer to use the calibration based on the TMK04 models instead of the differential models because the latter are computed at fixed [Fe/H] rather than [Z/H] which is the parameter that we estimate from the galaxy spectra in our sample (see Sec.~\ref{sec:age_met}).

Our method relies on the assumption that the BC03 models have a solar abundance ratio $\afe=0$. However this is strictly true only for metallicities close to solar. The BC03 models follow the abundance pattern of the Milky Way and thus correspond to significantly super-solar \afe~for metallicities $\log(Z/Z_\odot)\lesssim-0.5$. To gain insight into the potential bias, we compare the trend of the index ratio \mgtfe~with metallicity for BC03 SSP models with that for TMK04 models at \afe=0 and for $\alpha$-MILES models at \afe=0, after having normalized all models at the same \mgtfe~for solar metallicity. The offset in \mgtfe~between BC03 models and truly solar-scaled models should be proportional to the difference in \afe. Taking TMK04 as reference we would estimate BC03 SSP to have \afe~increasing from 0.1 to 0.2~dex for metallicities between [Z/H]$=-0.4$ and $-0.7$ and for ages older than 3~Gyr. Taking $\alpha$-MILES as reference we would estimate BC03 SSPs to have \afe~increasing from 0.2 and 0.5 for the same age and metallicity ranges. The estimated \afe~ would reduce to $0.1-0.15$~for younger ages. In these age and metallicity regimes our \afe~could be underestimated by this amount. The majority of our sample galaxies have stellar metallicities $\log(Z/Z_\odot)>-0.4$~and the scaling relations are not affected by this bias. Only the low-mass end of the \afe$-M_\ast$ relation for star-forming galaxies may be biased low: we estimate that their average \afe~could be underestimated by 0.1~dex at masses $\rm M_\ast\leq10^{9.5}M_\odot$. In any case, this is not a concern for the differential comparison between centrals and satellites, given the negligible differences in metallicity found in this regime.

We note that the relations between \afe~and stellar mass (halo mass), both globally and split into halo mass bins (stellar
mass bins), obtained adopting different calibrations between \Dmgfe~and \afe~are consistent with each other. We caution though 
that the overall \afe~range and the slope of the \afe$-M_\ast$ relation are sensitive to the adopted calibration: in
particular we obtain \afe~extending to higher values and a steeper slope with stellar mass adopting the
TMK04, \cite{TMJ11} or $\alpha$-MILES models rather than the differential models (the median \afe~of $10^{11.5}M_\odot$~galaxies varies between 0.3 and 0.45 dex among the various calibrations; see also Walcher et al 2015). However,
the comparison between central galaxies and satellites split into halo or stellar mass bins is remarkably
consistent. The conclusions presented in this paper are thus not affected by the chosen calibration.

\begin{figure}
\centerline{\includegraphics[width=9truecm]{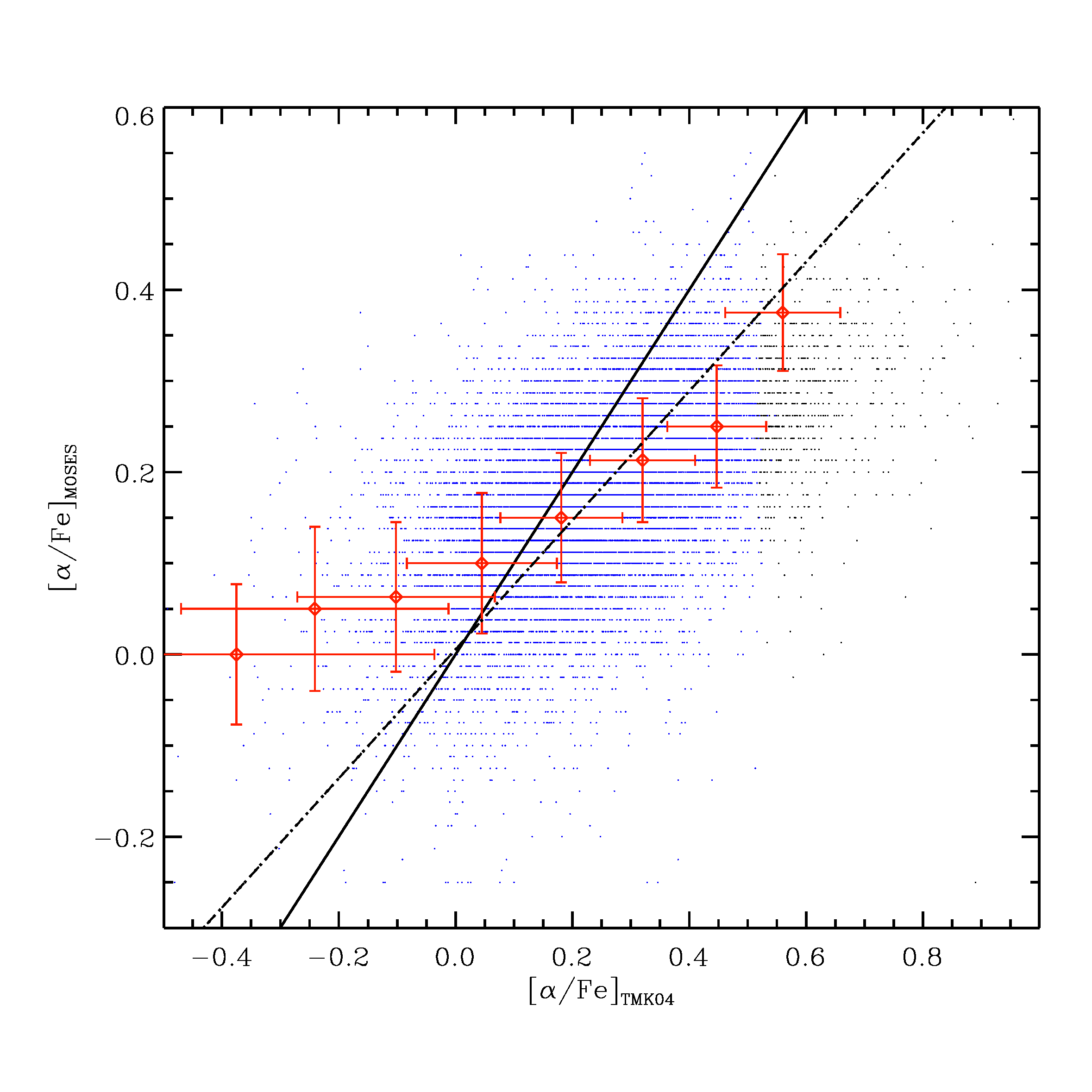}}
\caption{Comparison between the \afe~obtained by Thomas et al (2010) (\afe$_{MOSES}$) and those estimated in this work based on the Thomas et al (2003) and Thomas et al (2004) models (\afe$_{TMK04}$) for the 12896 galaxies in common with the MOSES sample (Schawinski et al 2007) of early-type galaxies (black dots; blue dots are galaxies for which no extrapolation of the relation \Dmgfe$-$\afe~ of the models was needed). Red circles indicate the median \afe$_{MOSES}$ in bins of \afe$_{TMK04}$ with error bars reflecting the median uncertainty on both parameters.The relation is flatter than the 1:1 relation (solid line), as indicated by the linear fit to \afe$_{MOSES}$ versus  \afe$_{TMK04}$ (dashed line).}\label{fig:compare_moses}
\end{figure}

As a further check, for the 12896 galaxies in common, we have compared our estimates of \afe~with those obtained by \cite{Thomas10} for the MOSES sample of morphologically-selected early-type galaxies \citep{schawinski07}. The comparison is shown in Fig.~\ref{fig:compare_moses}. There is a good correlation between the two estimates with a scatter of 0.08 (less than 10\% of the \afe~range), but with a slope of 0.38 in the sense that our estimates are higher.  

\section{Aperture effects}\label{A2}
The galaxies in our sample span a range in redshift and galaxies with higher stellar mass and/or in more massive halos tend to be observed out to higher redshift than lower mass galaxies or galaxies in smaller halos. Therefore the fixed-aperture SDSS spectra sample a different fraction of the galaxy light as a function of the galaxy mass and redshift. While the
overall trends of galaxy population properties with mass are not strongly affected by this aperture bias \citep[see also][for a discussion on aperture effects]{gallazzi05,gallazzi08}, we want to check whether the (small) differences between satellites in massive halos (Fig.\ref{fig:resid_mstar_sfr}) and centrals are affected by aperture bias. 

\begin{figure}
\centerline{\includegraphics[width=9truecm]{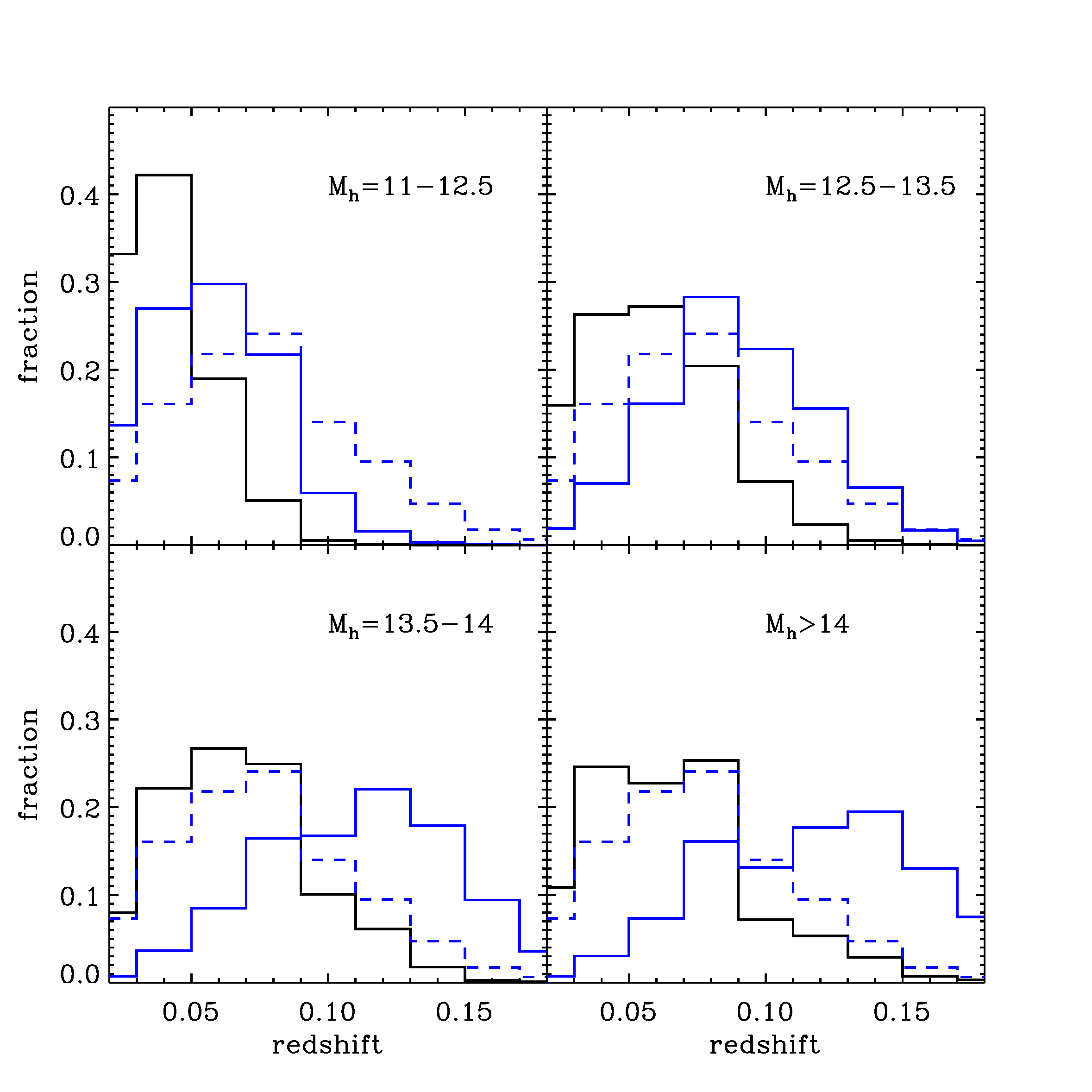}}
\caption{Distribution in redshift for satellites (black solid lines) and centrals (blue solid lines) in different halo mass bins. The distribution for central galaxies as a whole is reported in each panel with a dashed line.}\label{fig:zdistr_mhalo}
\end{figure}

We notice that the distribution in redshift is a function of halo mass for satellites and, even more strongly, for centrals (see Fig.\ref{fig:zdistr_mhalo}): galaxies in lower mass halos are detected out to $z<0.1$, while for higher mass halos there is an increasing fraction of galaxies detected out to higher $z$. The redshift distributions for satellites and for centrals in the same halo mass bin deviate more and more going from low- to high-mass halos. However, the distribution in redshift for satellites in halos more massive than $\rm \log M_h \geq 13.5$ is similar to that of the general central galaxies population. 
We repeat Fig.\ref{fig:resid_mstar_sfr} in narrow redshift ranges ($0.02-0.05; 0.05-0.07; 0.07-0.09; 0.09-0.12; >0.12$). We find that the trends observed for quiescent galaxies in each redshift bin, in particular the slight excess for satellites in massive halos with respect to centrals, are consistent with the ones observed for the sample in the full redshift range. For star-forming and green-valley galaxies the excess for massive satellites in massive halos observed marginalizing over redshift disappears at $z<0.05$. These results indicate that aperture bias is not driving the excess in age, metallicity and \afe~observed for satellites in massive halos. However some aperture effect might be present in low-redshift star-forming galaxies, possibly indicating that the outer parts, sampled only at high redshift, quench earlier than the central parts, as expected under the action of ram-pressure stripping.

\section{Trends of centrals with stellar and halo mass}\label{A3}
\begin{figure}
\centerline{\includegraphics[width=9truecm]{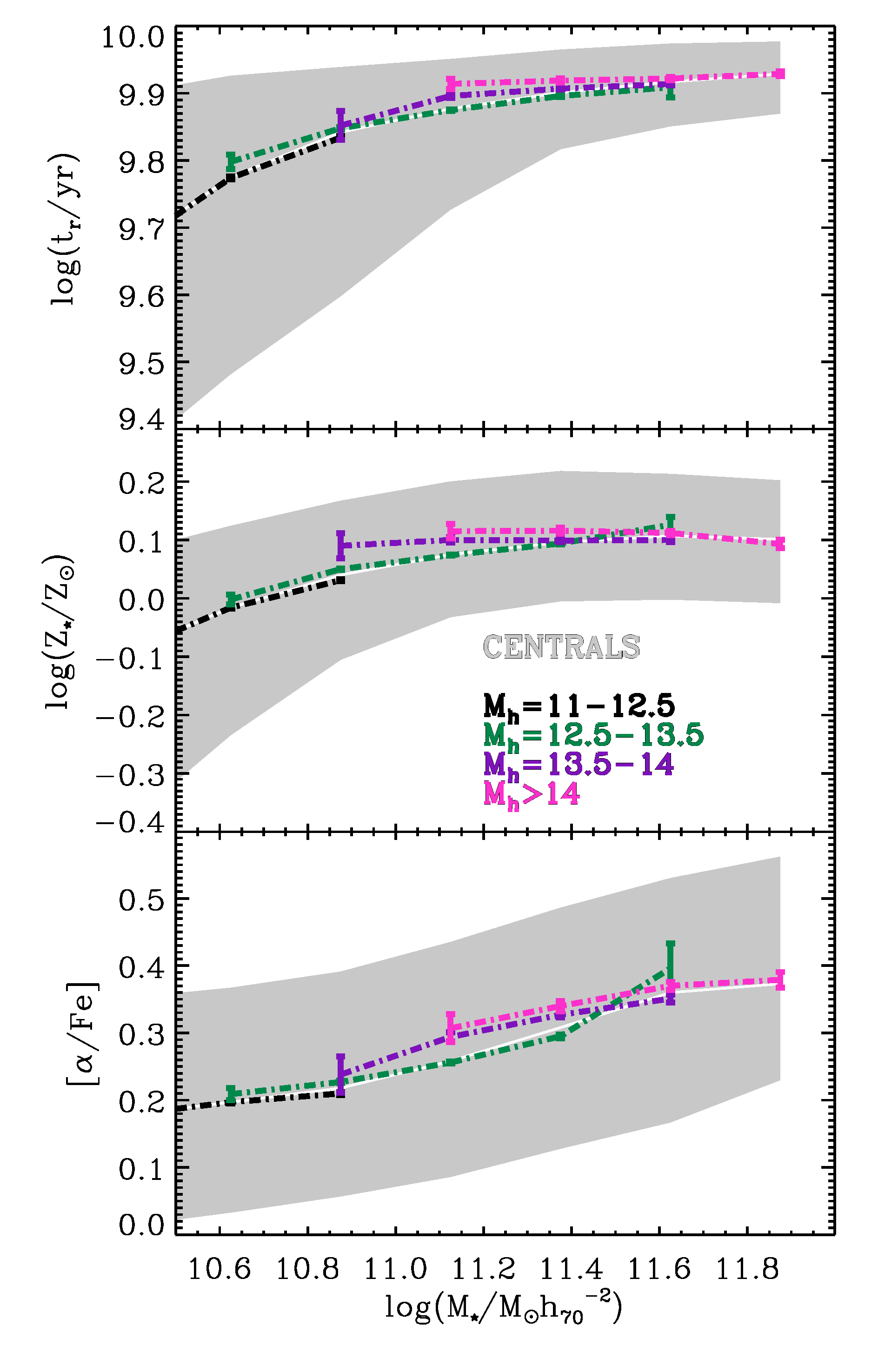}}
\caption{Light-weighted age (upper panel), stellar metallicity (middle panel), \afe~(bottom panel) as a function of stellar mass for central galaxies only. The white line and grey region in each panel indicate the median and $16-84$~interpercentile range of the whole central population, while the colored dot-dashed lines show the median trends for centrals in different bins of halo mass. The errorbars indicate the error on the median. With respect to Fig.\ref{fig:rel_mstar_mhalo} we focus on the high mass range to highlight the small trends of centrals with halo mass at fixed stellar mass.}\label{fig:rel_cen_zoom}
\end{figure}

In Fig.\ref{fig:rel_cen_zoom} we zoom onto the relations between age, stellar metallicity, \afe~and stellar mass for only central galaxies divided into bins of halo mass, as in Fig.\ref{fig:rel_mstar_mhalo}. This plot shows more clearly the trends discussed in Sec.\ref{sec:cen_mhalo}.

% Don't change these lines

\bsp
\label{lastpage}
\end{document}